\documentclass[sigconf,nonacm]{acmart}
\usepackage{algorithm}
\usepackage{algorithmic}

\usepackage{xcolor}
\renewcommand{\textcolor}[2]{#2}
\usepackage{xcolor}
\usepackage{subfig}
\usepackage{balance}

\AtBeginDocument{%
  }

\setcopyright{acmlicensed}
\copyrightyear{2025}
\acmYear{2025}
\acmDOI{XXXXXXX.XXXXXXX}
\acmConference[Conference acronym 'XX]{Make sure to enter the correct
  conference title from your rights confirmation email}{June 03--05,
  2018}{Woodstock, NY}
\acmISBN{978-1-4503-XXXX-X/2018/06}

\usepackage{enumitem}
\usepackage[normalem]{ulem}  

\usepackage{bm}
\begin{document}

\title[Scaled Block Vecchia Approximation for High-Dimensional Gaussian Process Emulation on GPUs]{Scaled Block Vecchia Approximation for High-Dimensional Gaussian Process Emulation on GPUs}


%
\author{Qilong Pan$^{1,5}$, Sameh Abdulah$^{1,5}$, Mustafa Abduljabbar$^{1,5}$, Hatem Ltaief$^{1,5}$, Andreas Herten$^{2,6}$, \\ Mathis Bode$^{2,7}$, Matthew Pratola$^{3,8}$, Arindam Fadikar$^{4,9}$, Marc G. Genton$^{1,5}$, \\ David E. Keyes$^{1,5}$, Ying Sun$^{1,5}$}
\vspace{4mm}
\affiliation{
\institution{
$^1$King Abdullah University of Science and Technology, Thuwal, KSA}  \city{}  \country{}}

\affiliation{\institution{$^2$Jülich Supercomputing Centre (JSC), Jülich, Germany}\city{}  \country{}}
\affiliation{\institution{$^3$Indiana University, Bloomington, IN, USA}\city{}  \country{}} 
\affiliation{\institution{$^4$Argonne National Laboratory, Lemont, IL, USA}\city{}  \country{}}
\affiliation{$^5$\textit {\{Firstname.Lastname\}@kaust.edu.sa}\,\,$^{6}$\textit {a.herten@fz-juelich.de}\,\,\\$^{7}$\textit {m.bode@fz-juelich.de}\,\,$^{8}$\textit {mpratola@iu.edu}\,\,$^{9}$\textit {afadikar@anl.gov} \city{}  \country{}}







\settopmatter{authorsperrow=1}
\renewcommand{\shortauthors}{Qilong Pan et al.}

\begin{abstract}
Emulating computationally intensive scientific simulations is crucial for enabling uncertainty quantification, optimization, and informed decision-making at scale. Gaussian Processes (GPs) offer a flexible and data-efficient foundation for statistical emulation, but their poor scalability limits applicability to large datasets. We introduce the Scaled Block Vecchia (SBV) algorithm for distributed GPU-based systems. SBV integrates the Scaled Vecchia approach for anisotropic input scaling with the Block Vecchia (BV) method to reduce computational and memory complexity while leveraging GPU acceleration techniques for efficient linear algebra operations. To the best of our knowledge, this is the first distributed implementation of any Vecchia-based GP variant. Our implementation employs MPI for inter-node parallelism and the MAGMA library for GPU-accelerated batched matrix computations. We demonstrate the scalability and efficiency of the proposed algorithm through experiments on synthetic and real-world workloads, including a 50M point simulation from a respiratory disease model. SBV achieves near-linear scalability on up to 512 A100 and GH200 GPUs, handles 2.56B points, and reduces energy use relative to exact GP solvers, establishing SBV as a scalable and energy-efficient framework for emulating large-scale scientific models on GPU-based distributed systems.
\end{abstract}

\keywords{Batched Linear Algebra, Energy-Efficient HPC, Gaussian Processes, GPU Computing, High-Dimensional Data, Scalable Emulation, Vecchia Approximation}

\maketitle
\section{Introduction}
Many modern scientific and engineering advances rely on the use
of computer models that use mathematical equations to represent physical systems and are implemented as high-fidelity simulations. These models allow researchers to explore phenomena that are impractical to observe experimentally, such as climate dynamics \cite{eyring2016overview, kochkov2024neural}, materials \cite{borvik2001computational,walters2018bayesian}, and cosmological structure formation \cite{springel2005cosmological, spurio2022cosmopower, lawrence2017mira}. Computer models can generate predictions that inform policy decisions, drive technological innovation, and facilitate fundamental discoveries by solving systems of differential equations or leveraging particle-based methods \cite{borvik2001computational, eyring2016overview, springel2005cosmological}. However, when exploring models with many input parameters, learning the relationship between inputs and outputs may require thousands of evaluations \cite{fiore2018road}. Such high computational costs often constrain their usage, motivating the adoption of statistical emulators to approximate the simulator's behavior at a fraction of the cost \cite{santner2003design,abdulah2024boosting}.

 


A common approach to building computer model emulators is to fit a Gaussian Process (GP) to data generated by running the simulator at a predefined set of input values (design points), thereby learning a mapping from the model input space to the model output space \cite{andrianakis2012effect}. GP-based emulators enhance interpretability and facilitate efficient analysis, particularly when simulating scenarios across a broad parameter space. For example, variable importance \cite{saltelli2010variance} and sequential optimization \cite{winkel2021sequential} may only be feasible using a cheap-to-evaluate GP emulator of the underlying expensive-to-evaluate simulator.  
\textcolor{red}{ While methods such as RBF-based models \cite{acosta1995radial, arora2023review} can offer flexible function approximation for such endeavors, they are fundamentally deterministic heuristics that lack marginal likelihood formulations, statistically grounded parameter estimation, and coherent uncertainty quantification. As a result, they are not well aligned with the probabilistic and inference-driven objectives of this work.  In contrast, GP modeling provides a fully probabilistic framework with principled uncertainty quantification and likelihood-based inference, which are essential when studying scalable approaches to inference and made possible by covariance approximations
such as Vecchia-based methods \cite{vecchia1988estimation, katzfuss2021general}.}

However, GPs face significant scalability challenges due to their $\mathcal{O}(n^3)$ computational cost and $\mathcal{O}(n^2)$ memory requirements, where $n$ is the number of data points. These costs arise from operations on a dense $n \times n$ covariance matrix that captures the correlations between data inputs ~\cite{abdulah2018exageostat}. For instance, modeling $200$K locations demands about $300$ GB of memory and an estimated $2.6$ PFLOP of computation, making standard GPs infeasible for large-scale applications without approximation methods. Thus, numerous studies have addressed the computational and memory challenges of large-scale GPs. Some efforts have pursued exact GP computation on High-Performance Computing (HPC) systems \cite{abdulah2018exageostat, abdulah2019geostatistical, abdulah2021accelerating}. For instance, the largest reported problem size was processed on ORNL's Frontier system, handling $27.24$M data points in mixed-precision across approximately $9$K nodes ($\sim$36K AMD GPUs) of the system \cite{abdulah2024boosting}. However, most of the literature has focused on approximation to alleviate the computational burden of GP modeling, including low-rank \cite{furrer2006covariance,bevilacqua2016covariance,shi2025decentralized}, tile low-rank  \cite{abdulah2018parallel,salvana2022parallel,mondal2023tile}, covariance tapering \cite{furrer2006covariance}, and Vecchia approximations \cite{katzfuss2021general,pan2024gpu}.

Among these approximation methods, Classic Vecchia (CV) approximation demonstrates superior advantages. In \cite{katzfuss2022scaled}, Katzfuss et al. introduced the Vecchia approximation for GP emulation in the context of computer experiments and proposed the Scaled Vecchia (SV) method to enhance approximation accuracy in high-dimensional settings. However, their study was limited to small-scale problems and relied entirely on CPU-based computations. In contrast, efforts have been made to accelerate the classic Vecchia algorithm on GPUs, as in \cite{pan2024gpu}, which subsequently introduced the Block Vecchia (BV) approach \cite{pan2025block} to further enhance computational speed while preserving approximation accuracy. Nevertheless, the use of CV approximation for high-dimensional GP emulation on GPUs remains unexplored. This gap presents an opportunity to develop a scalable GPU-based method that can efficiently address the computational and memory challenges posed by high-dimensional inputs.

In this work, we propose a novel algorithm, the Scaled Block Vecchia (SBV) approximation, for distributed GPU-based systems. SBV integrates the Block Vecchia (BV) algorithm from~\cite{pan2025block} with the Scaled Vecchia approach introduced in~\cite{katzfuss2022scaled}, while leveraging GPU acceleration techniques from~\cite{pan2024gpu}.
To the best of our knowledge, this work represents the first distributed implementation of any variant of the Vecchia-based GP algorithm. Leveraging the Message Passing Interface (MPI), we parallelize computations across multiple GPU-enabled nodes, achieving scalable performance well beyond that demonstrated in prior studies.

To validate the accuracy and robustness of our algorithm, we perform GP simulations across a range of parameter settings and benchmark its performance on the satellite drag dataset, a standard high-dimensional testbed for evaluating GP-based models~\cite{sun2019emulating}. As an application in computer model emulation, we apply our algorithm to \texttt{MetaRVM}, a compartmental simulation model designed for generic respiratory virus diseases~\cite{MetaRVM}. Experimental results demonstrate both high predictive accuracy and strong scalability, with successful runs on up to 512 A100/GH200 GPUs and problem sizes of up to $2.56$B data points. Additionally, our approach achieves a noticeable reduction in power consumption compared to the state-of-the-art exact GP framework on GPUs, i.e., \texttt{ExaGeoStat}~\cite{abdulah2018exageostat}.

The paper is organized as follows: Section 2 summarizes the main contributions of the study. Section 3 provides a brief overview of related work. Section 4 presents the necessary background. Section~5 details the proposed algorithm. Section 6 describes the simulation study and reports the results of the computer model application. Section 7 focuses on performance evaluation and analysis. Finally, conclusions are presented in Section 8.

\textcolor{red}{
\section{Contributions}
The key contributions of this work are as follows:

\begin{itemize}[leftmargin=*]

\item We propose the Scaled Block Vecchia (SBV) approximation, a novel GP algorithm that integrates anisotropic input scaling with block-based conditioning. While BV enables scalable inference for large spatial datasets in low-dimensional settings and SV improves approximation accuracy in high-dimensional inputs, the regime of high-dimensional and large-scale GP inference remains largely unresolved. SBV bridges this gap by combining the advantages of SV and BV, improving approximation accuracy while reducing both computational and memory complexity.

\item We develop efficient clustering and neighbor-selection algorithms, including Random Anchor Clustering (RAC) and a filtered block-wise KNN construction, that reduce the clustering and neighbor-search costs to approximately linear complexity. These algorithms make SBV practical for billion-point, high-dimensional datasets where standard clustering and nearest-neighbor construction would otherwise be computationally prohibitive.

\item We develop a distributed implementation of SBV for modern GPU-accelerated HPC systems. Our implementation employs MPI for inter-node parallelism and the MAGMA library~\cite{abdelfattah2021set, agullo2009numerical} for fine-grained batched GPU-accelerated linear algebra operations.

\item We benchmark SBV on synthetic datasets and the high-dimensional satellite drag dataset, demonstrating substantial improvements in runtime and predictive accuracy over state-of-the-art GP approximation methods.

\item We apply SBV to emulate the \texttt{MetaRVM} epidemiological simulation using $50$M samples and $10$-dimensional inputs, achieving high predictive accuracy and practical scalability.

\item We demonstrate both weak and strong scaling on the JURECA-DC GPU (AMD EPYC + A100) and JUPITER (GH200 Superchip) systems at Jülich Supercomputing Centre (JSC), with experiments scaling up to $512$ nodes and problem sizes reaching $2.56$B data points.

\item We perform a comprehensive energy analysis across GPU platforms, showing that SBV significantly reduces power consumption compared to exact GP frameworks such as \texttt{ExaGeoStat} while maintaining high throughput and accuracy.

\end{itemize}

}



\section{Related Work}

\textbf{GP emulators:} GPs are widely used for emulating high-fidelity simulators in scientific computing, with applications in cosmology~\cite{spurio2022cosmopower}, climate modeling~\cite{abdulah2024boosting}, and mechanism science~\cite{sun2019emulating}. 
Notable examples include emulation of N-body cosmological simulations~\cite{lawrence2017mira}, particle-material interactions~\cite{walters2018bayesian}, and epidemiological models of disease spread~\cite{fadikar2018calibrating}.  While recent advances in scalable GPs have improved performance, few approaches offer distributed, multi-GPU implementations. The proposed SBV algorithm directly addresses this gap by enabling efficient, scalable GP emulation on distributed, GPU-accelerated HPC systems.

\textbf{CV approximation:}  CV approximation ~\cite{vecchia1988estimation,katzfuss2021general} reduces the computational complexity of GPs by factorizing the joint distribution into a product of conditional distributions. 
Recent advancements have enhanced the effectiveness of the CV approximation by improving data-point ordering and neighbor selection~\cite{guinness2018permutation}. The BV approximation~\cite{pan2025block} builds on this by conditioning on blocks of observations rather than individual points, enabling batched linear algebra operations on GPUs and improving throughput and memory efficiency in high-performance environments~\cite{pan2024gpu}. More recently, the SV method was introduced to address challenges associated with high-dimensional inputs by applying anisotropic input scaling~\cite{katzfuss2022scaled}, which enables the model to capture directional-varying correlations. Our work extends these developments by integrating block-based conditioning and anisotropic scaling into a unified framework, i.e., SBV, and presenting a distributed implementation optimized for multi-GPU systems.

\textcolor{red}{
\textbf{Vecchia-Based Applications and Extensions}
Recent research has increasingly adopted the Vecchia approximation as a scalable computational primitive within broader probabilistic modeling frameworks. For example, Vecchia approximations have been incorporated into deep Gaussian processes to enable fully Bayesian posterior inference with near-linear scaling for large computer experiments~\cite{sauer2023vecchia}. In the context of latent Gaussian process models with non-Gaussian likelihoods, iterative methods have been developed to accelerate Vecchia-Laplace approximations using conjugate gradients and stochastic trace estimation~\cite{kundig2025iterative}. It has also been extended to scalable Gaussian process regression with variable selection through penalized likelihood formulations and mini-batch subsampling~\cite{cao2022scalable}, and adapted to large-scale Bayesian optimization by integrating approximate nearest-neighbor search and variance recalibration strategies~\cite{jimenez2023scalable}. Beyond regression settings, Vecchia likelihood approximations have been employed for computationally tractable inference in high-dimensional spatial max-stable models~\cite{huser2024vecchia}, and have recently enabled linear-cost estimation of multivariate normal probabilities via sparse inverse Cholesky structures~\cite{cao2025linear}.
}
\textcolor{red}{
Previous studies have shown that the Vecchia approximation is a versatile and scalable tool for probabilistic modeling in various statistical fields. In contrast, our work pursues a different goal. We focus on likelihood-based Gaussian process emulation for high-dimensional inputs, combining scaled and block conditioning strategies, distributed-memory parallelization using MPI, GPU-accelerated batched dense linear algebra, and energy-efficient analysis on modern GPU supercomputers with billion-scale datasets. To our knowledge, this is the first distributed GPU implementation of a Vecchia-based Gaussian process method that scales to billions of data points, advancing Vecchia approximations from methodological scalability to extreme-scale HPC through accelerators.
}

\section{Background}
This section provides background on GPs, the BV approximation, Kullback–Leibler (KL) divergence, and the scaled kernel function, which together form the foundation for our proposed extension for high-dimensional data. Table~\ref{tab:abbreviations} summarizes the abbreviations used throughout the paper to enhance clarity and readability.

\begin{table}[h]
    \centering
    \caption{List of abbreviations and their explanations.}
    \label{tab:abbreviations}
    \resizebox{\columnwidth}{!}{
    \begin{tabular}{ll}
        \toprule
        Abbreviation & Explanation \\ 
        \midrule
        GPs  & Gaussian Processes \\ 
 CV/SV/BV/SBV & Classic/Scaled/Block/Scaled Block Vecchia \\
        $n$/$n^*$ & Problem size in estimation/prediction \\
        $bs_{est}$/$bc_{est}$ ($bs/bc$) & Block (batched) size/count for estimation\\
        $bs_{pred}$/$bc_{pred}$ ($bs^*/bc^*$) & Block (batched) size/count for prediction \\
        $m_{\text{est}}$/$m_{\text{pred}}$ & \# nearest neighbors for estimation/prediction \\
        \textcolor{red}{
        $k_p$} & \textcolor{red}{\# clusters/anchors} \\
        \bottomrule
    \end{tabular}
    }
\end{table}

\subsection{Gaussian Processes (GPs)}
GPs provide a flexible framework for modeling and predicting functions across low- and high-dimensional input spaces. Let \( \bm{X} = [\bm{x}_1, \bm{x}_2, \dots, \bm{x}_n]^\top \), where each \(\bm{x}_i \in \mathbb{R}^d\), denote the \(n\) input points with corresponding observations \(\bm{y} = [y_1, y_2, \dots, y_n]^\top\). A GP with zero mean and kernel \(K_{\bm{\theta}}(\bm{x}_i, \bm{x}_j)\) is formulated as \(\bm{y} \sim \mathcal{N}(\bm 0, \Sigma_{\bm{\theta}})\) and its log-likelihood can be represented as shown in \cite{rasmussen2006gaussian}: 
\begin{equation}
\ell(\bm{\theta}; \bm{y}) = -\frac{n}{2} \log(2\pi) - \frac{1}{2} \log |\Sigma_{\bm{\theta}}| - \frac{1}{2} \bm{y}^\top \Sigma_{\bm{\theta}}^{-1} \bm{y},
\end{equation}
where \(|\cdot|\) means determinant, \(\Sigma_{\bm{\theta}} \in \mathbb{R}^{n \times n}\) has entries \([\Sigma_{\bm{\theta}}]_{ij} = K_{\bm{\theta}}(\bm{x}_i, \bm{x}_j)\) parameterized by \( \bm{\theta} \). For a set of \(n^*\) new input points \(\bm{X}^* = [\bm{x}_1^*, \bm{x}_2^*, \dots, \bm{x}_{n^*}^*]^\top\), we are interested in predicting their corresponding observations \(\bm{y}^* = [y_1^*, y_2^*, \dots, y_{n^*}^*]^\top\). The joint distribution of \(\bm{y}\) and \(\bm{y}^*\) is assumed to be:
\[
\begin{bmatrix}
\bm{y} \\
\bm{y}^*
\end{bmatrix}
\sim \mathcal{N} \left(
\bm{0},
\begin{bmatrix}
\Sigma_{\bm{\theta}} & \Sigma_{\bm{\theta},*} \\
\Sigma_{\bm{\theta},*}^\top & \Sigma_{\bm{\theta}}^*
\end{bmatrix}
\right),
\]
where \(\Sigma_{\bm{\theta},*} \in \mathbb{R}^{n \times n^*}\) contains the covariance matrix between \(\bm{X}\) and \(\bm{X}^*\), and \(\Sigma_{\bm{\theta}}^* \in \mathbb{R}^{n^* \times n^*}\) contains the covariances among \(\bm{X}^*\). The conditional distribution of \(\bm{y}^*\) given \(\bm{y}\) is:
\[
p(\bm{y}^* | \bm{y}) \sim \mathcal{N}(\mu^*, \Sigma^*),
\]
where the conditional mean \(\mu^*\) and covariance \(\Sigma^*\) are derived as
\(
\mu^* = \Sigma_{\bm{\theta},*}^\top \Sigma_{\bm{\theta}}^{-1} \bm{y},
\)
\(
\Sigma^* = \Sigma_{\bm{\theta}}^* - \Sigma_{\bm{\theta},*}^\top \Sigma_{\bm{\theta}}^{-1} \Sigma_{\bm{\theta},*}.
\) The GPs have computational complexity \(\mathcal{O}(n^3)\) and memory complexity \(\mathcal{O}(n^2)\), \textcolor{red}{making it infeasible on large scale problems \cite{williams2006gaussian, vecchia1988estimation}.}

\subsection{Block Vecchia Method}

The BV approximation \cite{pan2024gpu} reduces the computational and memory cost of GPs by partitioning/clustering the dataset into \(bc\) disjoint blocks, \(\{\bm{B}_1, \bm{B}_2, \dots, \bm{B}_{bc}\}\). After applying a permutation \(\zeta\) to order the blocks, the exact likelihood of \(\bm{y}\) becomes:
\[
p_{\bm{\theta}}(\bm{y}) = \prod_{i=1}^{bc} p_{\bm{\theta}}(\bm{y}_{\bm{B}_i^\zeta} | \bm{y}_{\bm{B}_1^\zeta}, \bm{y}_{\bm{B}_2^\zeta}, \dots, \bm{y}_{\bm{B}_{i-1}^\zeta}).
\]

\begin{figure}[htbp]
    \centering
    \includegraphics[width=0.85\linewidth, page=1]{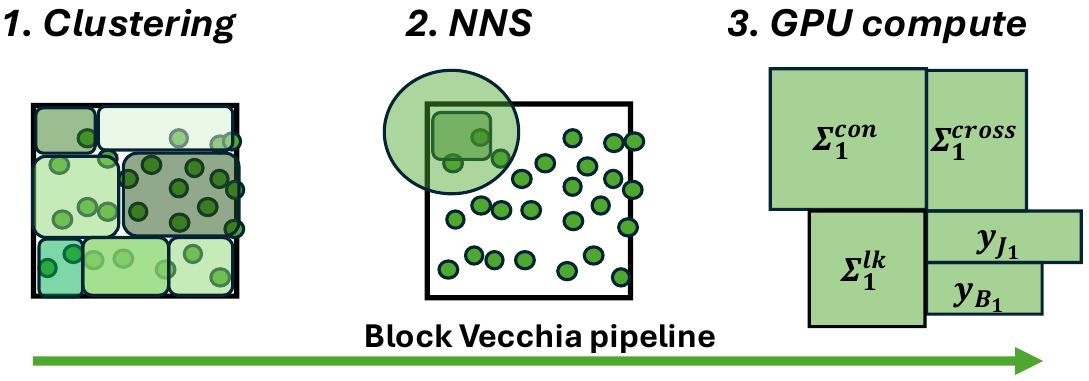}
\caption{
The BV algorithm: (1) disjoint clustering, (2) Nearest Neighbor Searching (NNS), and (3) batched GPU log-likelihoods computation.
}
\label{fig:bv-pipline}
\end{figure}

The BV method approximates each conditional distribution by replacing  each  vector with a subset of neighbors \(\bm{NN}(\bm{B}_{i}^\zeta)\) by:
\begin{equation}
\label{eq:mmsearchingBV}
p_{\bm{\theta}}(\bm{y}) \approx \prod_{i=1}^{bc} p_{\bm{\theta}}(\bm{y}_{\bm{B}_i^\zeta} | \bm{y}_{\bm{NN}(\bm{B}_i^\zeta)}),
\end{equation}
where the number of \(\bm{NN}(\bm{B}_{i}^\zeta)\) is \(m\) and \(m \ll n\). The prediction for \(\bm{y}^*\) under the BV framework approximates the conditional distribution \(p(\bm{y}^* | \bm{y})\) by the product of block-level conditional distributions. Given the new blocks \(\{\bm{B}_1^*, \dots, \bm{B}_{bc^*}^*\}\) on \(\bm{X}^*\), the predictive distribution is:
\begin{equation}
p_{\bm{\theta}}(\bm{y}^* | \bm{y}) \approx \prod_{j=1}^{bc^*} p_{\bm{\theta}}(\bm{y}_{\bm{B}_j^*} | \bm{y}_{\bm{NN}(\bm{B}_j^*)}),
\end{equation}
where \(\bm{NN}(\bm{B}_j^*)\) is the neighbors selected from the \(\bm{y}\). The BV method reduces the computational complexity of the exact GPs from \(\mathcal{O}(n^3)\) to \(\mathcal{O}(bc\cdot m^3)\) and memory complexity from \(\mathcal{O}(n^2)\) to \(\mathcal{O}(bc \cdot m^2)\).  Figure \ref{fig:bv-pipline} shows the computations involved in the BV algorithm \cite{pan2024gpu}, which includes three components, clustering \(\{\bm{B}_1, \bm{B}_2, \dots, \bm{B}_{bc}\}\), Nearest Neighbor Search (NNS) \(\bm{NN}(\bm{B}_{i}^\zeta)\), and batched GPU computing.

\subsection{Kullback-Leibler (KL) Divergence}
The KL divergence measures how one probability distribution diverges from a second probability distribution \cite{cover2006elements}. For two Gaussian distributions \(\mathcal{N}_0\) and \(\mathcal{N}_1\) with zero mean and covariance matrices \(\bm{\Sigma}_0\) and \(\bm{\Sigma}_1\), the KL divergence is \cite{murphy2012machine}:
\[
D_{KL}(\mathcal{N}_0 \| \mathcal{N}_1) = \frac{1}{2} \left\{ \text{tr}(\bm{\Sigma}_1^{-1} \bm{\Sigma}_0) - n + \log \frac{|\bm{\Sigma}_1|}{|\bm{\Sigma}_0|} \right\}.
\]

For Vecchia-based GPs, previous work~\cite{pan2025block, pan2024gpu} shows that the KL divergence simplifies to the difference between exact and approximate log-likelihoods evaluated at \(\bm{y} = \bm{0}\) as:
\begin{equation}
    \label{eq:kldivergence}
    D_{KL} = \ell_0(\bm{\theta}; \bm 0) - \ell_a(\bm{\theta}; \bm 0),
\end{equation}
where \(\ell_0(\bm{\theta}; \bm 0)\) is the exact log-likelihood and \(\ell_a(\bm{\theta}; \bm 0)\) is Vecchia-based approximated log-likelihood.

This work uses the KL divergence to quantify the discrepancy between the proposed SBV algorithm and posterior distributions from other GP methods.

\subsection{Scaled Kernel Function}

In high-dimensional spaces, covariance functions often require \emph{anisotropic scaling}, where each input dimension is scaled according to its relevance to the output \cite{katzfuss2022scaled}. This allows the model to reflect varying sensitivities across dimensions, thereby improving both accuracy and efficiency. The  scaled covariance function is defined as
\begin{equation}
    K_{\bm{\theta}}(\bm{x}_k, \bm{x}_{k'}) = f \left( \left( \sum_{i=1}^d \frac{(x_{ki} - x_{k'i})^2}{\bm{\beta}_i^2} \right)^{\frac{1}{2}}\right),
    \label{eq:scale}
\end{equation}
where \(x_{ki}\) and \(x_{k'i}\) denote the \(i\)-th components of \(\bm{x}_k\) and \(\bm{x}_{k'}\), and \(\bm{\beta} = (\beta_1, \dots, \beta_d)^\top\) contains the dimension-specific range (scaling) parameters. \(f(\cdot)\) is a kernel function, such as the Mat\'ern kernel:
\begin{equation}
    f(r) = \sigma^2 \frac{2^{1-\nu}}{\Gamma(\nu)} r^\nu \mathcal{K}_\nu(r) + \sigma_0^2.
    \label{eq:maternkernel}
\end{equation}
Here, \(r\) is the scaled distance and \(\bm{\theta} = (\sigma^2, \bm{\beta}, \nu, \sigma_0)^\top\) denotes the kernel parameters: \(\sigma^2\) (process variance), \(\sigma_0^2\) (nugget), and \(\nu > 0\) (smoothness). \(\Gamma(\cdot)\) is the gamma function, and \(\mathcal{K}_\nu(\cdot)\) is the modified Bessel function of the second kind of order \(\nu\). Additional valid covariance classes are detailed in \cite{genton2001classes}. We use the scaled form in Equation~\eqref{eq:scale} to capture directional relevance, useful in emulation problems where only a subset of input dimensions significantly affect the output.

\section{Distributed Scaled Block Vecchia Algorithm}

\begin{figure}[htbp]
    \centering
    \includegraphics[width=1\linewidth, page=1]{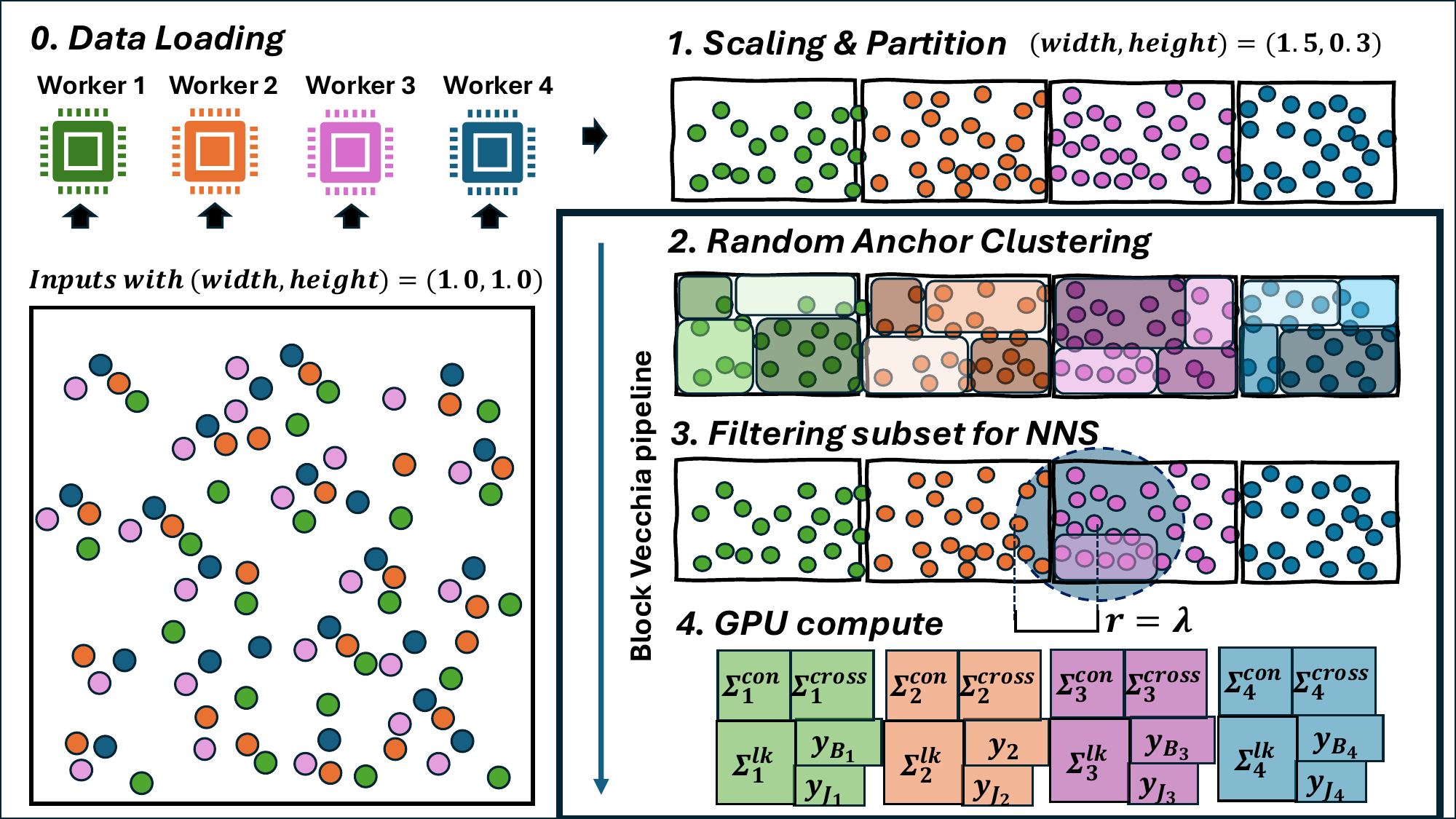}
\caption{Distributed SBV pipeline.
\textcolor{black}{The pipeline for the distributed SBV algorithm: (0) parallel data loading; (1) anisotropic scaling and partitioning; (2) Random Anchor Clustering (RAC) to form disjoint spatial blocks; (3) neighborhood filtering for NNS within a radius $r = \lambda$; and (4) batched GPU computation of blockwise conditional log-likelihoods.
}}
\vspace{-4mm}
    \label{fig:pipeline}
\end{figure}

This section presents the proposed distributed Scaled Block Vecchia (SBV) algorithm.
The algorithm comprises four key components: scaling and partitioning, Random Anchor Clustering (RAC), filtered subset selection for NNS, and batched GPU computation to optimize the log-likelihood. The preprocessing steps, i.e., scaling and partitioning, RAC, and filtered NNS, are performed once on the CPU to prepare the desired data structure.  The resulting structured data is then transferred to the GPU, enabling efficient execution of hundreds of parameter optimization iterations without redundant data movement. The distributed SBV algorithm is detailed in Algorithm \ref{alg:main}, and the pipeline is illustrated in Figure \ref{fig:pipeline}, where
Steps (2), (3), and (4) are the Block Vecchia (BV) algorithm pipeline.

\begin{algorithm}[htbp]
\caption{Distributed SBV approximation algorithm}
\begin{algorithmic}[1]
\label{alg:main}
\STATE \textbf{Input:} Data $\{\bm{x}_{i}, y_i\}_{i=1}^n$, dimension $d$, total block count $K$,  nearest neighbors $m_{\text{est}}$, workers $P$, covariance function $K_\theta$, scaling parameters $\{\beta_i\}_{i=1}^d$.
\STATE \textbf{Output:} Approximate the log-likelihood $\ell$.

\STATE \textbf{Step 0: Data Loading $X^{org}_p$}

\STATE Each worker load local points $X^{org}_p$, $p=1,2,\ldots, P$.

\STATE \textbf{Step 1: Partitioning and Scaling $X^{org}_p$}
\STATE $X_{1:P} = \mathcal{PS}(X^{org}_{1:P})$

\STATE \textbf{Step 2: Random Anchor Clustering $\bm B_p$}
\STATE $\{\bm B_{p,i};p=1,2,\dots, P; i = 1,2, \dots, k_p\} = \mathcal{C}(X_{1:P})$
\STATE Randomly reorder these blocks $B_{p,i}$;
\STATE \textbf{Step 3: NNS $\bm J_{p}$}
\STATE $\{\bm J_{p,i};p=1,2,\dots, P; i = 1,2, \dots, k_p\} = \mathcal{V}(\bm B_{1:P}, m_{\text{est}})$

\STATE \textbf{Step 4: Batched Log-Likelihood Calculation $\log \mathcal{L}_p$}
\FOR{each block $\bm B_{p}, p = 1, 2, \ldots, P$ in Parallel}
    \STATE $\ell_{p} = \ell_{llh}(\bm B_{p}, \bm J_{p}, \bm{y}_{p, \bm B}, \bm{y}_{p, \bm J})$, where $\bm{y}_{p, \bm B}$, $\bm{y}_{p,\bm J}$ are the observations associated with $\bm B_{p}$ and $\bm J_{p}$ respectively. 
\ENDFOR
\STATE \textbf{Step 5: Reduction of Log-Likelihood Across Workers}
\STATE Use MPI\_Allreduce to sum $ \ell_p$ from each worker and obtain the approximated log-likelihood $\ell$.

\STATE \textbf{Return:} The approximated log-likelihood $\ell$.

\end{algorithmic}
\end{algorithm}

\subsection{Data Partitioning and Scaling  }
\textcolor{black}{Each worker \( p \) begins by loading its assigned data subset \( \bm{X}_p \). Firstly, the points are partitioned into several segments along their most relevant dimension, as shown in line 4 of Algorithm \ref{alg:scalingpartition}. Here, the most relevant dimension is denoted by \( d_{max} \), and \texttt{MPI\_Alltoall} is used to redistribute the dataset \( \bm{X}_{1:P} \), where \( x_{p,i,d_{max}} \) refers to the \( d_{max} \)-th component of the \( i \)-th data point on worker \( p \). Secondly, input dimensions are rescaled using anisotropic parameters \( \bm{\beta} = [\beta_1, \ldots, \beta_d] \). This scaling transforms each dimension \( x_{ij} := x^{org}_{ij}/\beta_j \), ensuring that the relevance of different dimensions is appropriately reweighted (see line 7 in Algorithm \ref{alg:scalingpartition}).}

\textcolor{black}{Figure~\ref{fig:pipeline} gives an example of partitioning within the unit square $[0, 1]^2$. During data loading, each color denotes the portion assigned to a specific worker; for example, the green points are assigned to Worker 1. In the scaling and partitioning step, the input space is partitioned into four segments along the most relevant dimension, with \( \beta_2 = 0.3 \). After that, the input space is rescaled from the unit square to the rectangle $(0, 1.5) \times (0, 0.3)$, indicating that the first dimension (width) is more relevant to the response than the second (height). This rescaling and partitioning strategy ensures balanced and uniform data distribution across workers in the algorithm, enhancing computational efficiency while preserving accuracy.}

\begin{algorithm}[htbp]
\caption{Partitioning and Scaling $\mathcal{PS}(\bm{X}^{org}_{1:P})$}
\label{alg:scalingpartition}
\begin{algorithmic}[1]
\STATE \textbf{Input:}  Complete dataset $\bm{X}^{org}_{1:P}$.
\STATE \textbf{Output:} Scaled and redistributed 
$\bm{X}_{1:P}$. 

Relevant dimension $d_{max} =\arg \max_i \{1/\beta_i, i=1,2, \dots, d\}.$ 

\FOR{each worker $p$ in parallel}
    \STATE \texttt{MPI\_Alltoall}, 
    send $\bm x^{org}_{p,i}$ to $q$th node if 
    $\text{int} (\bm x^{org}_{p,i, d_{max}} \cdot P) = q$.
\ENDFOR

\FOR{each worker $p$ in parallel}
    \STATE \textcolor{black}{Dimension scaling, $\bm{X}_{p,j} := \bm{X}^{org}_{p,j} / \beta_j$, $j = 1, 2, \ldots, d$.}
    
\ENDFOR

\STATE \textbf{Return:} Redistributed and scaled 
$\bm{X}_{1:P}$

\end{algorithmic}
\end{algorithm}

\subsection{Random Anchor Clustering (RAC)}

\textcolor{black}{Next, small blocks $\bm B_p$ are created, defined as \(\bm B_{p}:= \{\bm B_{p,1}, \ldots, \bm B_{p,k_p}\} \), for each partition by RAC independently.} The RAC method is detailed in Algorithm \ref{alg:rac}. Here, we replace K-means clustering in block Vecchia GPs \cite{pan2025block} with RAC to reduce the computational cost incurred by K-means for large datasets while maintaining comparable approximation accuracy.
The RAC algorithm begins by randomly selecting $k_p$ anchors (centers) for each worker $p=1,2,\dots, P$ in parallel. These anchors serve as the centers of blocks. Subsequently, each $\bm x_{p,i}$ is assigned to the $j$th block based on their distance, as shown in line 5 of Algorithm \ref{alg:rac},
thus constructing clusters, $\bm B_p:=(\bm B_{p,1}, \bm B_{p,2}, \ldots, \bm B_{p,k_p})$ with $p=1,2,\dots, P$. The RAC method does not involve any communication between workers, considering that the size of clusters is typically much smaller than the overall dataset. For example, a single cluster may contain only 100 locations, whereas the entire dataset comprises 10 million locations. \textcolor{black}{ In Step 2 of Figure~\ref{fig:pipeline}, each worker creates 4 blocks in their data partition.}

\begin{algorithm}[htbp]
\caption{RAC algorithm $\mathcal{C}(\bm{X}_{1:P})$}
\label{alg:rac}
\begin{algorithmic}[1]
\STATE \textbf{Input:} $\bm{X}_{1:P}$ full dataset.
\STATE \textbf{Output:} $\bm{B}_{1:P}$ set of cluster for each node. 

\FOR{each worker $p$ in parallel}
    \STATE Randomly choose $k_p$ local centers $C_p=\{\bm{c}_{p,i} \mid i = 1, 2, \dots, k_p\}$ in $\bm{X}_p$, and initialize $\bm{B}_{p,j} = \{\bm c_{p,i}\}.$
    \STATE
    Assign $\bm x_{p,i}$  to $j$th cluster $\bm{B}_{p,j}$, where $j= \arg\min_{j} \|\bm{x}_{p,i} - 
    \bm{c}_{p,j}\|^2, j=1,2,\dots, k_q$.
\ENDFOR
\STATE \textbf{Return:} Clusters $\bm{B}_{1:P}$

\end{algorithmic}
\end{algorithm}

\subsection{Filtered $m$-Nearest Neighbor Search (NNS)}

With block sets $\bm B_{1:P}$, the next goal is to find the exact $m$ nearest neighbors for all block centers in high-dimensional space, 
where search areas are varied for each query center in the Vecchia-based GPs, see Equation \eqref{eq:mmsearchingBV}. Existing methods \cite{NNsurvey, NNconstraint, clarkson2006nearest} for exact $m$-NNS with query-specific search areas face limitations, especially for large datasets. 

To address these issues, we adopt the filtered $m$-NNS on each block \( \bm B_{p,i} \), detailed in Algorithm~\ref{alg:nn}. This approach selects a small filtered subset (a circle of radius $\lambda$) for brute-force $m$-NNS to obtain exact results while reducing computational cost. In Algorithms~\ref{alg:main} and~\ref{alg:nn}, $\mathcal{V}$ denotes the filtered $m$-NNS function, which begins by computing a Monte Carlo-based distance threshold. This threshold $\lambda$ is calculated using:

\begin{equation}
\label{eq:distancethreashold}
    \lambda = \left( \alpha \frac{m\zeta}{n}\right)^{1/d}, \quad \zeta = 
    \begin{cases}
    \frac{\Gamma\left( \frac{d}{2} + 1 \right)}{\pi^{\frac{d}{2}}} &  d\text{ is even} \\
    \frac{2 \pi^{\frac{d-1}{2}} \Gamma\left( \frac{d+1}{2} \right)}{\Gamma(d+1)} & d\text{ is odd}
    \end{cases}.
\end{equation}
Here, $n$ is the total number of points, $m$ is the number of nearest neighbors, $d$ is the dimensionality, and $\alpha$ is an expansion factor to account for irregularities in point distribution (e.g., $\alpha = 100$ implies the candidate set is expected to be 100 times larger than $m$).
Then, we prepare the small filtered subset in two steps: {\em coarser} and {\em finer} preparation. In the {\em coarser} preparation, each worker redistributes candidate clusters to the corresponding worker if the distance between two centers is within the calculated threshold using \texttt{MPI\_Alltoall}. The complementary coarser candidate set is denoted as $\bm B_{p}^{cand}$ (lines 10-13 in Algorithm \ref{alg:nn}).  In the {\em finer} partition, each worker independently selects finer candidates $\bm S_{p,i}$ for the $i$th cluster, i.e., $\bm x_{k}$ is added to the finer candidate set $\bm S_{p,i}$ if the distance between the $i$th local center and the data point $\bm x_{k}$ is within the threshold $\lambda$ (lines 15-22 in Algorithm \ref{alg:nn}).
Finally, each local center $\bm c_i$ applies brute-force search for its $m$ nearest neighbors $\bm J_{p,i}$ from the finer candidates set $\bm S_{p,i}$. The computational cost of brute search in $\bm S_{p,i}$ is negligible compared to $\mathcal{O}(n)$ due to $\alpha m \ll n$ and total block count $K \ll n$, thus filtered $m$-NNS achieves computational complexity $\mathcal{O}(\alpha m + K)$ and has a simple implementation. 

\begin{figure}[htbp]
    \centering
    \includegraphics[width=0.8\linewidth, page=2]{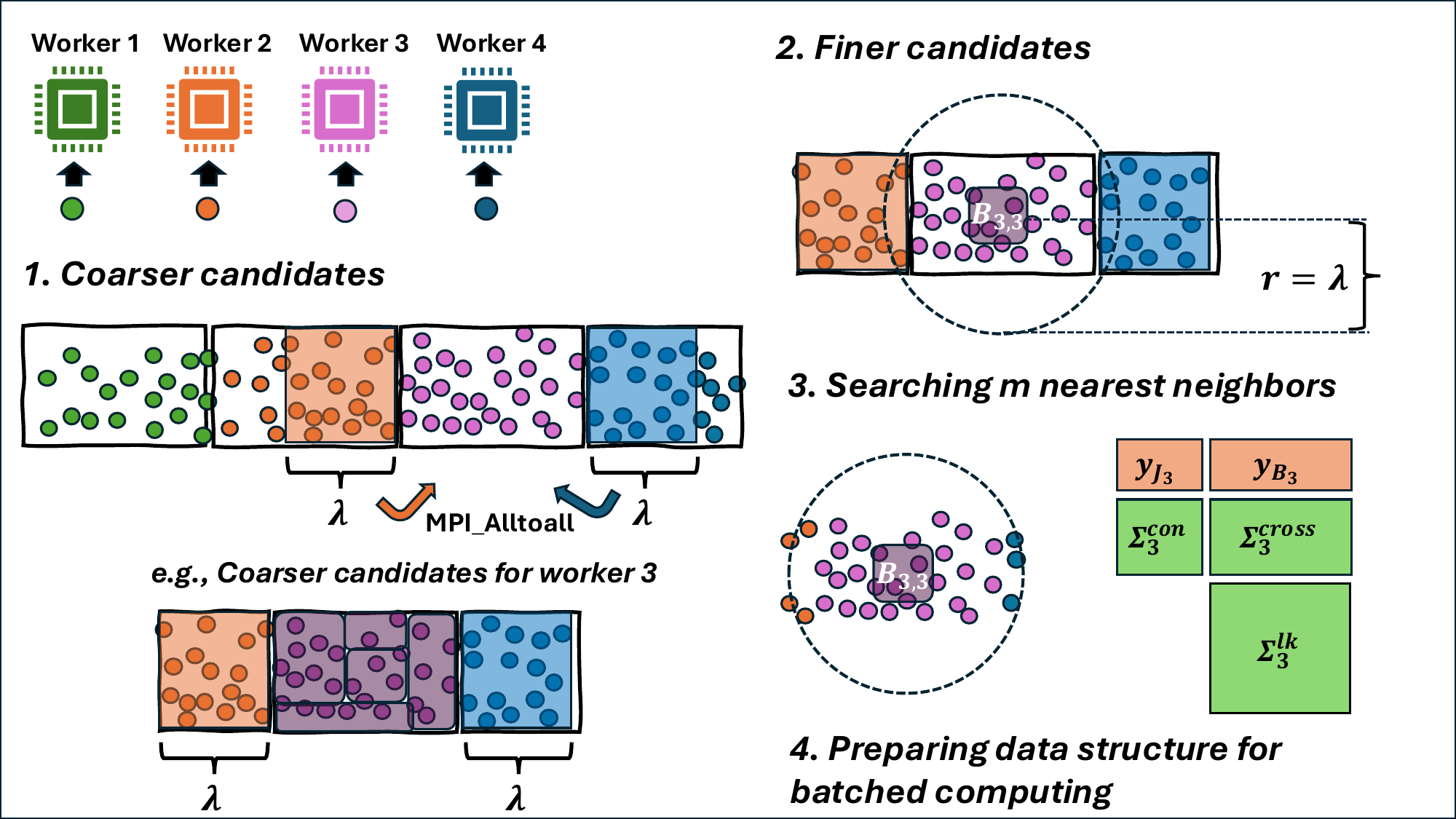}
\caption{
Filtered NNS algorithm pipeline. \textcolor{black}{ (1) workers expand local partitions by radius $\lambda$ and exchange boundary data; (2) finer candidates are selected within $r = \lambda$ around each block; (3) $m$ nearest neighbors are identified; (4) data and covariance matrices are generated.}
}
\vspace{-4mm}
    \label{fig:filterednns}
\end{figure}

Figure \ref{fig:filterednns} illustrates the filtered m-NNS using an example of four workers. In the coarser candidate preparation, \texttt{MPI\_Alltoall} is utilized to distribute related blocks (within distance threshold $\lambda$) to the $3$rd worker,  forming the coarse candidates $\bm B_3^{cand}$. Subsequently, the $3$rd block $\bm B_{3,3}$ refines candidates for the brute $m$-NNS (within distance threshold $\lambda$). Finally, blocks, their nearest neighbors, and their observations are arranged in a contiguous manner for batched GPU computation.

\begin{algorithm}[htbp]
\caption{NNS algorithm based on filtered subset $\mathcal{V} (\bm B_{p}, m)$}
\label{alg:nn}
\begin{algorithmic}[1]
\STATE \textbf{Input:} $\bm B_{p}$ cluster set, nearest neighbors $m$, expansion factor $\alpha=100$ (default).
\STATE \textbf{Output:} Nearest neighbor set $\bm J_{1:P}$

\STATE \textbf{Step1: Distance Threshold and Centers}

\STATE Calculate distance threshold, $\lambda$ according to Equation (\ref{eq:distancethreashold}).

\FOR{each block $\bm B_{p,i}$ in $\bm B_{p}$}

\STATE Update centers $\bm C_i = \{\bm c_{p,i}\}$, where $\bm c_{p,i} =  \frac{1}{k_{p,i}}\sum_{i=1}^{k_{p,i}}\bm x_{p,i}$ with $\bm x_{p,i} \in \bm B_{p,i}.$
\ENDFOR
\STATE \texttt{MPI\_Allgather}, gather all centers $\bm C_{1:P}$ to each worker.
\STATE \textbf{Step 2: Prepare Coarser Candidates}

\FOR{each worker $p$ in parallel}
    \STATE \texttt{MPI\_Alltoall}, redistribute $\bm B_{p,i}$ to $q$th node if  $\| \bm c_{p,i} - \bm c_{q,j}\|^2 \leq \lambda$ where $j$ exists.
    \STATE Received clusters form the candidate set $\bm B^{cand}_p$ for worker $p$.
\ENDFOR

\STATE \textbf{Step 3: Prepare Finer Candidates for $NN$}
    \FOR{local block center $\bm{c}_i$ in $\{\bm{c}_i\}_{i=1}^{k_p}$}
        \FOR{candidate block center $\bm{c}_j $ in $\bm B^{cand}_p$}
            \IF{The order $\bm c_j$ not exceeds order $\bm c_i$}
                \FOR{each point $\bm x_{k}$ in $\bm B^{cand}_{p,j}$}
                    \STATE Calculate the distance $d(\bm{c}_i, \bm{x}_k)$.
                    \textcolor{black}{If $d(\bm{c}_i, \bm{x}_k) < \lambda$, add $\bm{x}_k$ to the candidate set $\bm S_{p,i}$ for block $\bm B_{p,i}$.}
                \ENDFOR
            \ENDIF
        \ENDFOR
        \STATE Searching $m$ nearest neighbors for $i$th block, i.e., $\bm J_{p,i} = NNS(\bm c_i, \bm S_{p,i}, m)$.
    \ENDFOR
\STATE \textbf{Return:} Nearest neighbor set $\{\bm J_{p},  p = 1,2,\ldots, P\}.$

\end{algorithmic}
\end{algorithm}

\subsection{Batched Log-likelihood Computations}

\textcolor{red}{

While MPI enables scalable distributed computation across nodes, GPU acceleration delivers high single-node throughput for the SBV algorithm's log-likelihood computations. These computationally intensive phases involve batched dense factorizations and batched matrix multiplications resulting from block conditioning. These operations are entirely offloaded to the GPU using the MAGMA batched linear algebra library. Each conditioning block in the SBV algorithm is treated as an independent batch element, which exposes fine-grained parallelism across available accelerators. Batched Cholesky factorizations and GEMM updates are executed concurrently to ensure high device occupancy. Block sizes are chosen to balance kernel launch overhead, arithmetic intensity, and memory reuse. All GPU kernels operate in double precision (FP64) to maintain numerical stability and consistency with likelihood-based Gaussian process inference. Future work will investigate the extension to lower- or mixed-precision formats.

The SBV implementation is designed to efficiently utilize the GPU memory hierarchy. After partition construction on the CPU, all associated matrices are transferred to the high-bandwidth memory (HBM) and remain resident throughout the likelihood evaluation phase. The MAGMA library is used for batched linear algebra operations~\cite{tomov2009magma}. Conditioning blocks are reused across multiple batched operations, thereby increasing arithmetic intensity and reducing redundant memory accesses. The batched linear algebra kernels in SBV demonstrate high arithmetic intensity as a result of repeated reuse of conditioning matrices within each partition. This characteristic makes the workload well-suited for GPU acceleration.

Host-to-device transfers are performed only during the initial data staging following partition construction. No intermediate CPU–GPU synchronization is necessary during the batched factorization and matrix update phases. Device-to-host communication is restricted to the final likelihood values and gradient quantities. Consequently, host–device transfers do not occur within the primary iterative GPU kernels and account for only a small portion of the total runtime. Maintaining data residency on the GPU throughout the computational phase minimizes PCIe/NVLink traffic. The total host-to-device transfer volume per node is proportional to the local partition size and occurs once per likelihood evaluation. 

}

To explain this in detail, once data preprocessing is completed (Algorithm~4), the clusters $\bm B_{1:P}$, their corresponding nearest-neighbor sets $\bm J_{1:P}$, and the associated observations are transferred to the GPU global memory for subsequent log-likelihood optimization. Algorithm \ref{alg:batchedllh} outlines the computation of a single log-likelihood iteration, which must be evaluated multiple times until the prespecified optimization configuration is reached.
Here, the batched operations execute for log-likelihood computation. This involves constructing local covariance matrices \( \bm{\Sigma}^{lk}, \bm{\Sigma}^{cross} \) and \( \bm{\Sigma}^{con} \) using the scaled kernel \( K_{\bm{\theta}} \). Subsequently, the GPU data structure is constructed in Step 4 in Figure \ref{fig:pipeline}. Following, batched Cholesky decompositions (\texttt{POTRF}) is utilized to factorize neighbor covariance matrices, \( \bm{\Sigma}^{con} \). Conditional mean and covariance updates are computed via batched triangular solver (\texttt{TRSM}) and matrix multiplications (\texttt{GEMM}), while batched triangular solver (\texttt{TRSV}) and determinant calculations derive per-block log-likelihoods \( \ell_{p,i} \). Each node aggregates its local computations \( \ell_p = \sum_{i=1}^{k_p} \ell_{p,i} \), and a global reduction across all nodes yields the final log-likelihood \( \ell = \sum_{p=1}^P \ell_p \) using \texttt{MPI\_Allreduce}.

\begin{algorithm}[htbp]
\caption{Batched Block Vecchia $\ell_{llh}(\bm B_{p}, \bm J_{p}, \bm{y}_{p, B}, \bm{y}_{p, J})$}
\label{alg:batchedllh}
\begin{algorithmic}[1]
\STATE \textbf{Input:} In $p$th node, $\bm B_{p}$ cluster set, $\bm J_{p}$ nearest neighbor set, $\bm{y}_{p, B}$ and $\bm{y}_{p, J}$ observation sets.
\STATE \textbf{Output:} log-likelihood $\ell_p$ 
\STATE \textbf{Step 1: Conditional Mean and Covariance Update}
\STATE $\bm \Sigma^{lk} \gets \text{batched}$ $ K_{\bm\theta}(\bm B_p, \bm B_p)$.
\STATE $\bm \Sigma^{con} \gets \text{batched}$ $ K_{\bm\theta}(\bm J_p, \bm J_p)$.
\STATE $\bm \Sigma^{cross} \gets \text{batched}$ $ K_{\bm\theta}(\bm J_p, \bm B_p)$.
    \STATE $ \bm L \gets \text{batched}POTRF(\bm \Sigma^{con})$ 
    \STATE $ \bm \Sigma^{'cross} \gets \text{batched}TRSM(\bm L, \bm \Sigma^{cross})$
    \STATE $ \bm y'_{p, J} \gets \text{batched}TRSV(\bm L, \bm y_{p,j})$
    \STATE $ \bm \Sigma^{cor} \gets \text{batched}GEMM (transpose(\bm \Sigma^{'cross}), \bm \Sigma^{'cross})$
    \STATE $ \bm \mu^{cor} \gets \text{batched}GEMV ( transpose(\bm \Sigma^{'cross}), \bm y'_{p,J})$
\STATE  $\bm\Sigma^{new} \gets \bm\Sigma^{con} - \bm \Sigma^{cor}$ 
\STATE  $\bm\mu^{new} \gets \bm \mu^{cor}$

\STATE \textbf{Step 2: GPs calculations}
    \STATE $ \bm L' \gets \text{batched}POTRF(\bm\Sigma^{new})$
    \STATE $ \bm v \gets \text{batched}TRSV(\bm L', \bm y_{p, B} - \bm\mu^{new})$
    \STATE $ \bm u \gets \text{batched}DotProduct \left( transpose(\bm v), \bm v \right)$
    \STATE $ \bm d  \gets 2\times \log(determinant(\bm L'))$
    \STATE $\ell_p \gets - \frac{1}{2}\bm 1^T\left( \bm u  + \bm d  \right)$
    \RETURN log-likelihood $\ell_p$
\end{algorithmic}
\end{algorithm}

\vspace{-4mm}
\subsection{GP Prediction}

During the prediction stage, the pipeline follows the estimation process, with the key difference that, in Step (2) of Algorithm \ref{alg:batchedllh}, GP calculations are replaced by conditional simulations. Specifically, the variance vector,  $\bm{\sigma} = \left( \sigma_1, \sigma_2, \dots, \sigma_{n^*} \right)^\top$ is subtracted from the diagonal elements of the covariance matrices $\bm{\Sigma}^{new} $, and $\bm \mu^{new}$ serves as the predicted values, $({y}_{*1}, {y}_{*1}, \dots, {y}_{*n^*})$. {\color {black} 
Next, conditional simulations are performed by drawing 1000 samples from the normal distribution \( \mathcal{N}(y_{*j}, \sigma_j) \)}. Then, the sample mean $ \tilde{\mu}_j $ and variance $ \tilde{\sigma}_j^2 $ are computed, and the 95\% confidence interval is given by, $
(\tilde{\mu}_j - z_{\alpha/2} \tilde{\sigma}_j, \tilde{\mu}_j + z_{\alpha/2} \tilde{\sigma}_j),$ where $ \alpha = 0.05, j=1,2,\dots, n^*$, which follows the same as \cite{pan2025block}.

Overall, the workflow minimizes inter-node communication and computations by scaling and partitioning, RAC, and filtered $m$-NNS, while GPU-batched kernels exploit fine-grained parallelism for covariance and log-likelihood computations. Moreover, this algorithm incorporates anisotropic scaling for high-dimensional inputs to ensure accurate emulation for computer models. This framework enables scalable and efficient evaluation of log-likelihoods for large-scale GPs in high-dimensional settings.

\section{Complexity Analysis}
Herein, we analyze the memory and computational complexity of SV and SBV, focusing on the GPU stage, which dominates due to repeated iterations, unlike the one-time CPU preprocessing.

For SV, the memory footprint is $\mathcal{O}(nm^2 + nm)$, primarily due to the storage of $n$ small covariance matrices $\bm{\Sigma}_{1:n}$ and corresponding conditioning vectors $\bm{y}_{1:n}$. In contrast, SBV partitions the data into $bc$ blocks. Each block stores three covariance matrices, i.e., $\bm{\Sigma}_i^{\text{lk}}$, $\bm{\Sigma}_i^{\text{cross}}$, and $\bm{\Sigma}_i^{\text{con}}$, and two observation vectors: $\bm{y}_{J_i}$ for the conditioning set and $\bm{y}_{B_i}$ for the block itself. Assuming an average block size of $bs = n / bc$, the memory complexity for each SBV component is: $\mathcal{O}(n \cdot bs)$ for $\bm{\Sigma}_i^{\text{lk}}$, $\mathcal{O}(mn)$ for $\bm{\Sigma}_i^{\text{cross}}$, $\mathcal{O}(bc \cdot m^2)$ for $\bm{\Sigma}_i^{\text{con}}$, $\mathcal{O}(bc \cdot m)$ for $\bm{y}_{J_i}$, and $\mathcal{O}(n)$ for $\bm{y}_{B_i}$. Therefore, the total memory complexity is $\mathcal{O}(nm^2 + nm)$ for SV and $\mathcal{O}(n \cdot bs + mn + bc \cdot m^2)$ for SBV.

The dominant computational costs in SBV arise from Cholesky factorizations and matrix-matrix multiplications. The Cholesky operations contribute $\mathcal{O}(bc \cdot bs^3)$ for block matrices and $\mathcal{O}(bc \cdot m^3)$ for the conditioning sets. Matrix-matrix multiplications add a complexity of $\mathcal{O}(mn \cdot bs)$. Thus, the total computational complexity for SBV is $\mathcal{O}(n \cdot bs^2 + bc \cdot m^3 + mn \cdot bs)$, compared to $\mathcal{O}(n m^3)$ for SV, which performs a Cholesky factorization for each data point.

For high-dimensional input spaces, we recommend setting the number of nearest neighbors to four times the block size, i.e., $m = 4 \times bs$. Under this setting, the complexities are simplified and summarized in Table~\ref{tab:complexity}. SBV achieves lower memory and computational costs than SV, especially when leveraging GPU-batched operations for large-scale GP models.

\vspace{-1mm}

\begin{table}[htbp]
    \centering
    \caption{Complexity analysis for SV and SBV under $m=4 \times bs$.}
    \label{tab:complexity}
    \renewcommand{\arraystretch}{1.0}
    \begin{tabular}{@{} l|c|c @{}}
        \toprule
        \textbf{Method} 
        & \textbf{Memory} & \textbf{Computational} \\
        \midrule
        SV  & $\mathcal{O}(n m^2)$ & $\mathcal{O}(n m^3)$ \\
        SBV & $\mathcal{O}(n m)$   & $\mathcal{O}(n m^2)$ \\
        \bottomrule
    \end{tabular}
\end{table}

\vspace{-4mm}

\section{Simulation and Benchmark Evaluation}
This section begins with a synthetic GP simulation using the Mat\'ern kernel in high-dimensional input space, highlighting the critical role of scaling and clustering. We then apply the proposed SBV algorithm to the satellite drag benchmark dataset to demonstrate its effectiveness and accuracy in emulating complex computer models.

\subsection{Synthetic Data Simulation}

The simulation framework follows the design outlined in \cite{katzfuss2022scaled}, ensuring consistency for comparative analysis. The synthetic dataset is generated from a zero-mean GP with a Mat\'ern kernel ($\nu = 3.5$) using Equation~\eqref{eq:maternkernel}. The input space is, $\bm{x} \in [0, 1]^{10}$.
The response is modeled as $y(\mathbf{x}) \sim \mathcal{N}(\bm{0}, K_{\bm{\theta}}(\mathbf{x}, \mathbf{x}'))$, where the parameter vector is given by $\bm{\theta} = (\sigma^2, \tau^2, \bm{\beta})$. The true parameters are set as $(\sigma^2, \tau^2) = (1.0, 0)$, with range parameters $\beta_1 = \beta_2 = 0.05$ for the relevant dimensions and $\beta_3 = \beta_4 = \dots = \beta_{10} = 5$ for the irrelevant ones. We evaluate four Vecchia-based GPs on this synthetic dataset: CV \cite{vecchia1988estimation}, BV
\cite{pan2025block}, SV \cite{katzfuss2022scaled}, and our proposed SBV.
The choice of $bs_{est}=bs_{pred} = 10$ in BV and SBV reflects a balance between computational efficiency and predictive accuracy. Model performance is assessed using both the KL divergence defined in Equation~\eqref{eq:kldivergence} and the Mean Squared Prediction Error (MSPE). To isolate approximation error and avoid conflating it with parameter estimation error, the true parameters $(\sigma^2, \tau^2, \bm{\beta})$ are directly supplied to all four models.

\begin{figure}[htbp]
    \centering
    \subfloat[KL divergence for fitting.]{\includegraphics[width=0.5\linewidth]{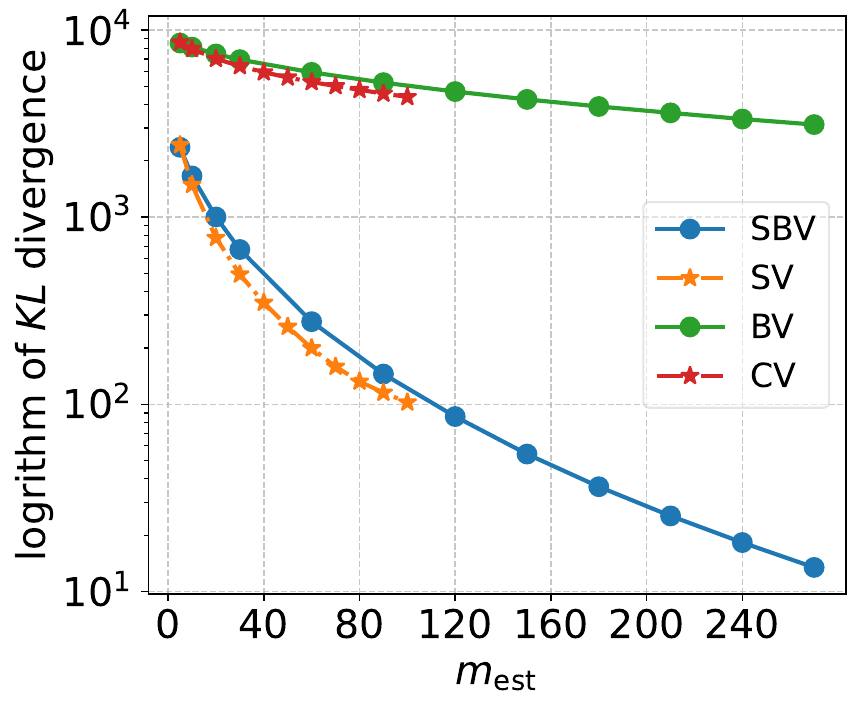}
    \label{fig:maternsimu1}}
    \subfloat[Prediction accuracy (MSPE).]{\includegraphics[width=0.5\linewidth]{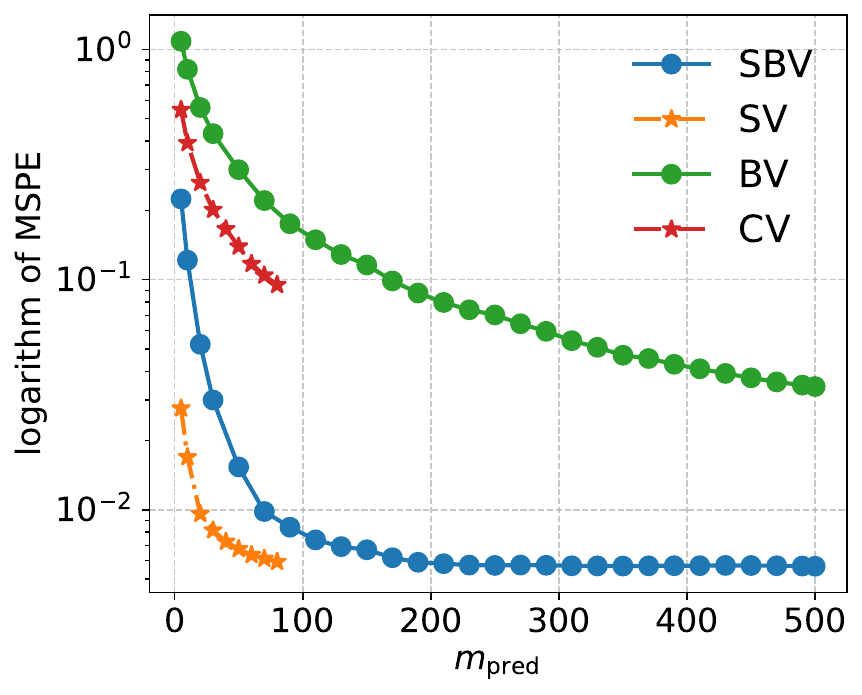}
    \label{fig:maternsimu2}}
    \\
    \subfloat[block size ($bs$) and SBV accuracy.]
    {\includegraphics[width=0.5\linewidth]{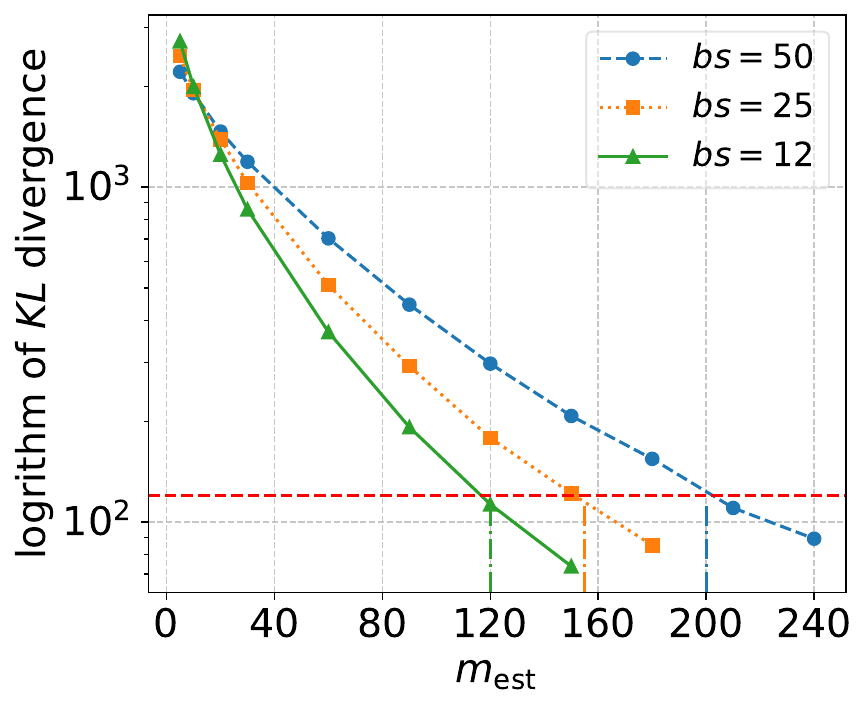}
    \label{fig:maternsimu3}}
    \subfloat[Relative error for RAC.]
    {\includegraphics[width=0.5\linewidth]{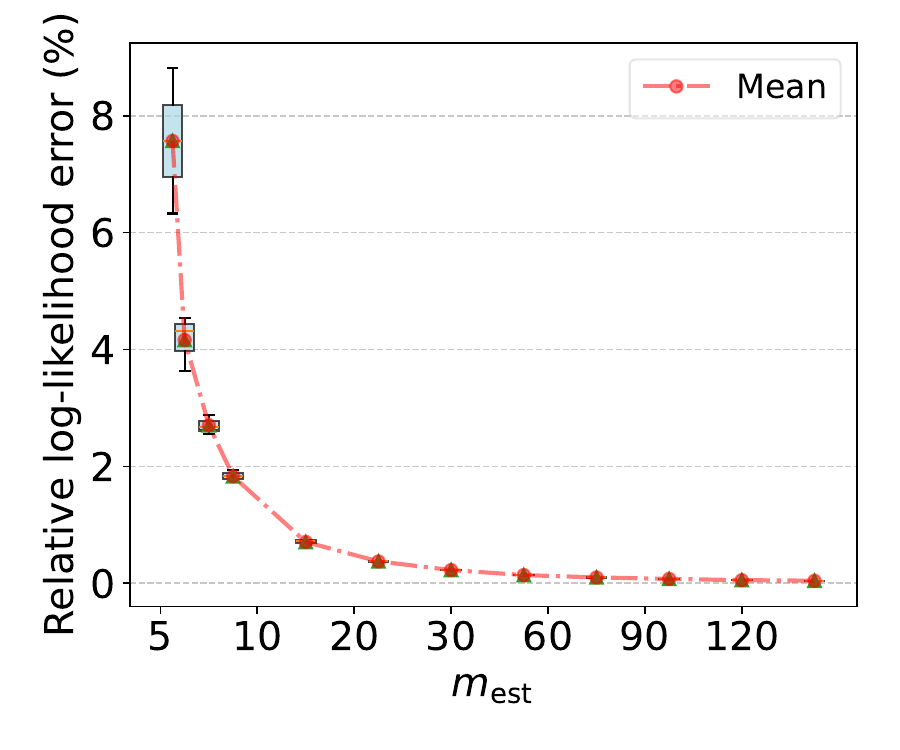}
    \label{fig:kmeans-vs-rac}}
    \caption{Comparison of Vecchia-based GP methods on model fitting and prediction accuracy, i.e., Classic Vecchia (CV), Block Vecchia (BV), Scaled Vecchia (SV), Scaled Block Vecchia (SBV). Subfigures (a) and (b) evaluate the GP variants fitting (KL divergence) and prediction separately, showing that scaling and block enable better accuracy; subfigure (c) Investigate SBV with different block sizes (bs), showing that a larger $m$ and a smaller block size yield the best fitting accuracy; \textcolor{red}{subfigure (d) Clarifying that the proposed RAC has comparable and robust log-likelihood with the K-means clustering and has the advantage of computationally linear complexity.}}
    \label{fig:maternsimu}
\end{figure}

The results are presented in Figure~\ref{fig:maternsimu}. In Figure~\ref{fig:maternsimu1}, the proposed SBV method achieves the lowest KL divergence, followed by SV, BV, and CV, highlighting the advantage of combining input scaling and clustering over the CV approach. The performance gains of SBV arise from its ability to address input anisotropy through joint scaling and to increase neighborhood coverage through clustering. 
Figure~\ref{fig:maternsimu2} shows a similar trend regarding predictive accuracy, where SBV again yields the lowest MSPE, followed by SV, BV, and CV. These results are consistent with the findings on the KL divergence and further validate the benefits of incorporating input scaling and structured clustering. In Figure~\ref{fig:maternsimu3}, we examine the impact of block size on KL divergence. Smaller blocks ($bs_{est} = 12$) yield lower KL divergence than larger ones ($bs_{est} = 50$), indicating improved approximation accuracy with finer partitioning. However, larger blocks impose heavier computational loads, thereby enabling effective GPU utilization. Therefore, the choice of block size $bs$ introduces a trade-off between accuracy and computational cost. 
\textcolor{red}{Figure~\ref{fig:kmeans-vs-rac} illustrates the relative error and its variance of the log-likelihood computed using two clustering methods: K-means and the proposed RAC approach. To assess the robustness of RAC with respect to anchor randomness, we perform five independent random anchor selections and report both the mean error and its variance. The results show that both the approximation error and its variability decrease as the conditioning size $m$ increases, eventually becoming nearly negligible. This trend confirms the stability of RAC for sufficiently large $m$. Furthermore, larger conditioning sizes are typically required in large-scale and high-dimensional applications to maintain high approximation accuracy.}

\subsection{Evaluation on Satellite Drag Benchmark}

The satellite drag dataset in \cite{sun2019emulating} is a widely used benchmark for evaluating GP-based models in high-dimensional settings \cite{sun2019emulating, katzfuss2022scaled}. It is generated from a high-fidelity simulator that models atmospheric drag coefficients in low Earth orbit (LEO). The dataset comprises $2$M simulation runs for each of six primary atmospheric species: atomic oxygen ($O$), molecular oxygen ($O_2$), atomic nitrogen ($N$), molecular nitrogen ($N_2$), helium ($He$), and atomic hydrogen ($H$). The total drag coefficient is computed as a weighted average of these species. Each species-specific dataset has an 8-dimensional input space: relative velocity, surface temperature, atmospheric temperature, yaw angle, pitch angle, and two accommodation coefficients.

Following the experimental design in \cite{katzfuss2022scaled}, we split the dataset into 90\% for training and 10\% for testing and performed 10-fold cross-validation for robust evaluation. We adopt the same Mat\'ern kernel with nugget effects, and smoothness parameter $\nu = 3.5$ as in \cite{katzfuss2022scaled}. Model performance is assessed using the Root Mean Squared Percentage Error (RMSPE) and the estimated kernel parameters. {\color {black}The models compared in this study are mentioned in Table \ref{tab:benchmark-models}}.

\begin{table}[htbp]
    \centering
    \caption{Vecchia-based GP configurations on the satellite drag dataset, showing block sizes ($bs_{\text{est}}$, $bs_{\text{pred}}$) and neighbor counts ($m_{\text{est}}$, $m_{\text{pred}}$) for estimation and prediction.}
    \label{tab:benchmark-models}
    \renewcommand{\arraystretch}{1.0}
    \begin{tabular}{l|ccccccc}
        \toprule
        Model & SV & SBV$_1$ & SBV$_2$ & SBV$_3$ & SBV$_4$ & SBV$_5$ & SBV$_6$ \\
        \midrule
        $bs_{\text{est}}$  & 1   & 100 & 100 & 100 & 100 & 100 & 100 \\
        $bs_{\text{pred}}$ & 1   & 5   & 5   & 5   & 5   & 5   & 5   \\
        $m_{\text{est}}$   & 50  & 200 & 200 & 200 & 400 & 400 & 400 \\
        $m_{\text{pred}}$  & 140 & 200 & 400 & 600 & 200 & 400 & 600 \\
        \bottomrule
    \end{tabular}
\end{table}

For model fitting, SV uses $m_{\text{est}} = 50$ nearest neighbors on a 50K-sample subset, following the configuration in \cite{katzfuss2022scaled}, which reported only marginal gains in predictive accuracy beyond this setting relative to the increased computational cost. In contrast, SBV leverages the full dataset with larger neighbor counts ($m_{\text{est}} = 200$ or $400$) while maintaining a consistent average block size of $bs_{\text{est}} = 100$ across all configurations. Due to the high memory and computational demands of SV, fitting the full dataset on a single GPU is infeasible, making SBV a more practical and scalable alternative.

\begin{figure}[htbp]
    \centering
    \includegraphics[width=1.0\linewidth]{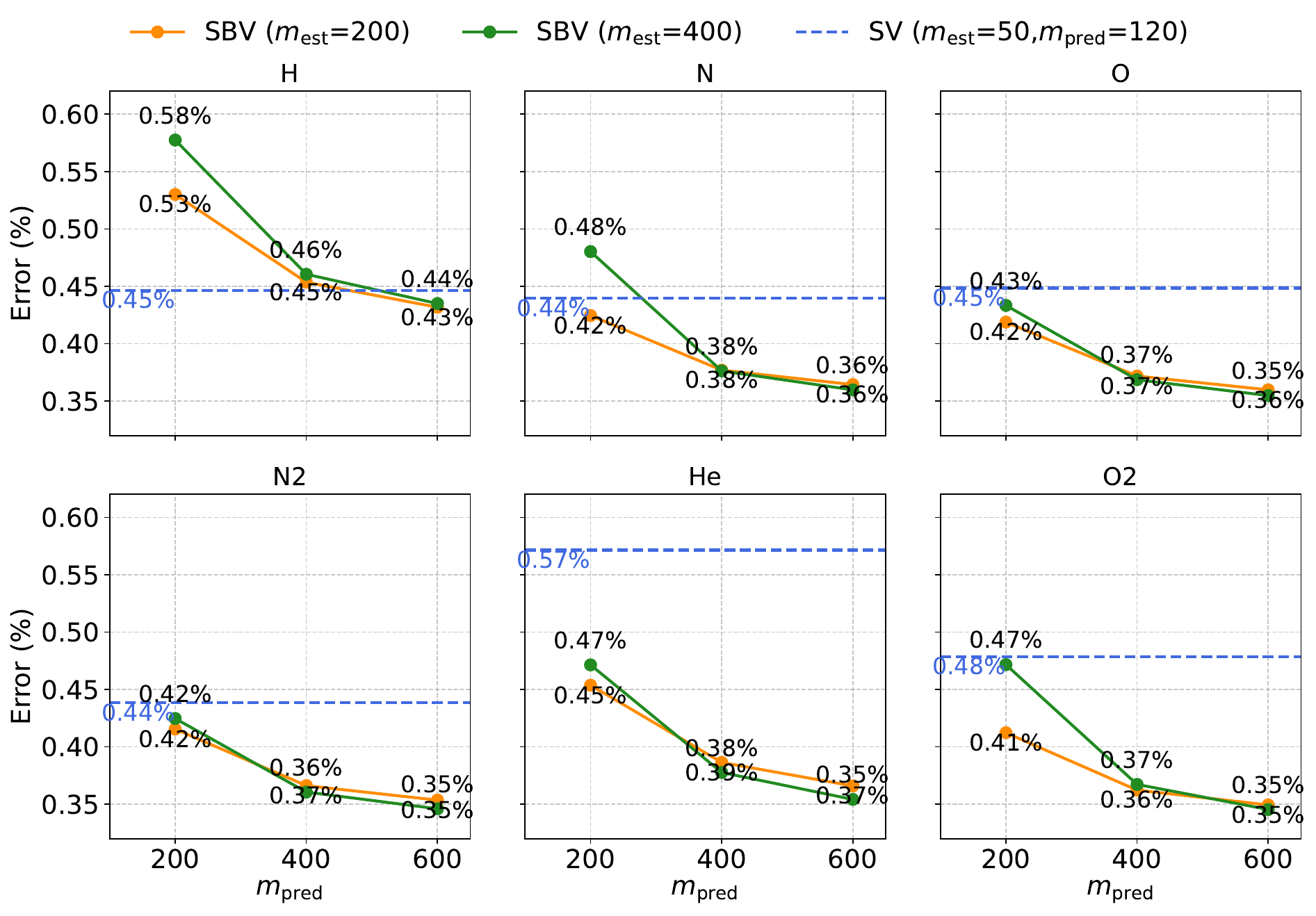}
    \caption{\textcolor{black}{Root Mean Squared Percentage Error (RMSPE)} of Vecchia-based GP models on six outputs from the satellite drag benchmark.}
    \label{fig:rmspe-benchmark}
\end{figure}

\begin{figure}[htbp]
    \centering
    \includegraphics[width=1.0\linewidth]{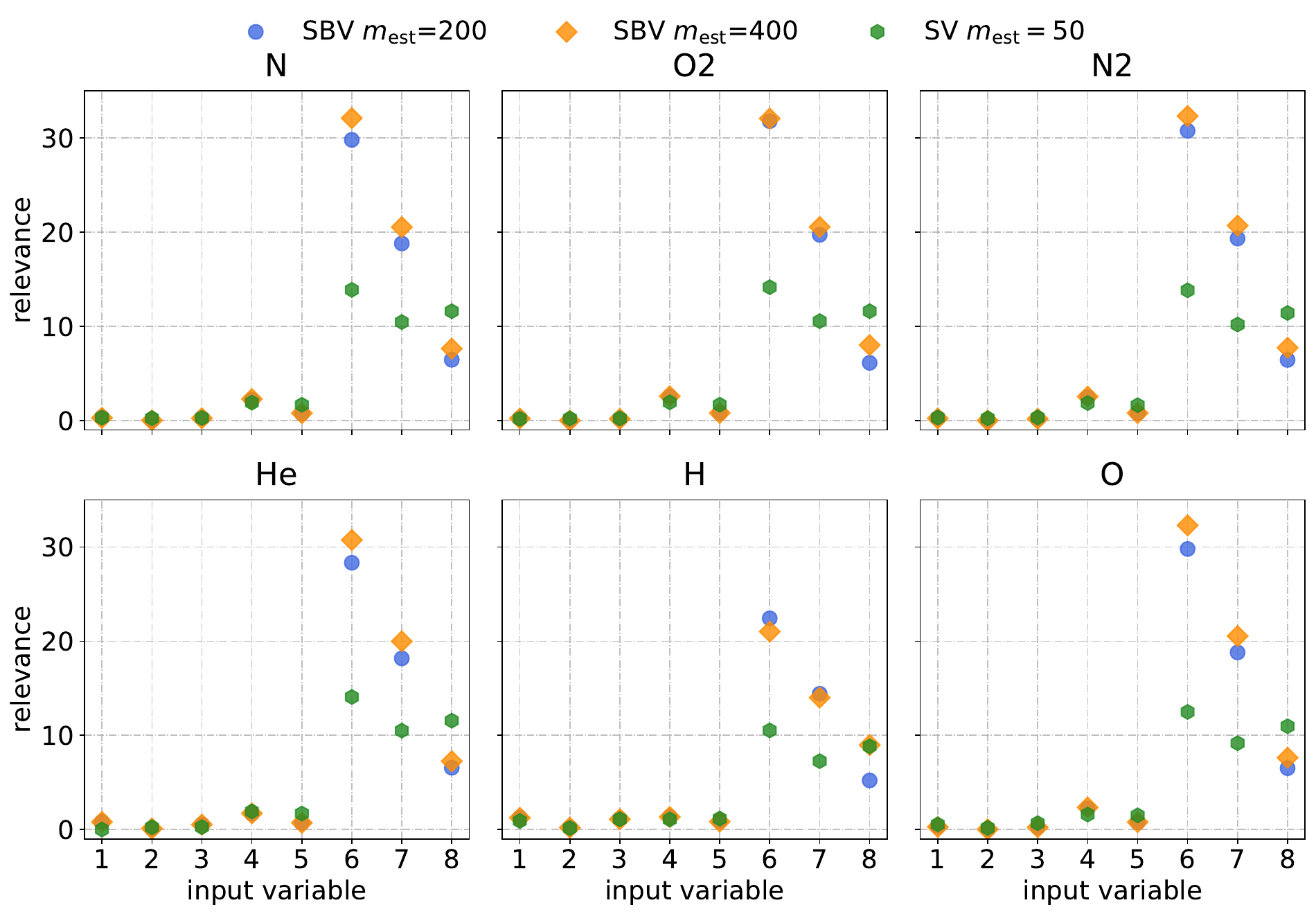}
    \caption{Estimated input relevance ($1/\beta_i$) for SV and SBV on the satellite drag benchmark. \vspace{-4mm}}
    \label{fig:params-benchmark}
\end{figure}

Figure~\ref{fig:rmspe-benchmark} presents the RMSPE results for Vecchia-based GP models on the benchmark dataset. SBV consistently achieves the lowest RMSPE, demonstrating the effectiveness of combining input scaling with clustering. Increasing the number of prediction neighbors $m_{\text{pred}}$ improves accuracy; for example, SBV with $m_{\text{pred}} = 400$ reduces RMSPE by approximately 15\% on average compared to $m_{\text{pred}} = 200$. 
To better understand the improved accuracy achieved by SBV, Figure~\ref{fig:params-benchmark} compares the estimated inverse lengthscales ($1/\beta_i$) for SV and SBV. Both methods identify the final three input dimensions as most relevant while treating the others as less influential. 
However, the parameter estimates from SBV differ significantly from those from SV, underscoring the importance of using a larger number of nearest neighbors for accurate GP fitting. From a computational and memory perspective, SV faces challenges in handling large neighbor counts because substantial increases in cost and memory usage limit efficient GPU utilization. In contrast, SBV remains scalable and efficient under these conditions.

\vspace{-3mm}
\subsection{\texttt{MetaRVM} Emulator}
\texttt{MetaRVM}~\cite{MetaRVM} is an R package that simulates the spread of respiratory viruses using a graph-based probabilistic model. By dividing a total population into sub-populations such as zones, age groups, or races, it models disease dynamics within and between these groups. This structure enables the simulation of interactions and disease spread within and between subpopulations. In healthcare, such models are crucial for understanding transmission patterns, evaluating intervention strategies, and informing public health policies. They help predict outbreak trajectories, assess the potential impact of vaccinations, and optimize resource allocation, ultimately aiding in the control and prevention of respiratory infections. The parameters in \texttt{MetaRVM} simulator are listed in Table \ref{tab:metarvm-params}. To understand the relationship between the input parameters and the accumulated number of hospitalizations (output) over 100 days in one population, we randomly choose 50M sets of input variables and obtain the accumulated hospitalizations using \texttt{MetaRVM} simulator. The 50M outputs are then divided into 90\% for fitting SBV GPs with Mat\'ern kernel at $\nu=3.5$ and 10\% is used for prediction, where the RMSPE is reported. The input is also scaled into $[0, 1]$, and the output is normalized with a mean of 1 to avoid the abnormal values in RMSPE. In the SBV, we set $bs_{\text{est}} = 100, bs_{\text{pred}} = 25$ and investigate the accuracy as increasing the number of nearest neighbors.

\begin{table}[h!]
\centering
\caption{The \texttt{MetaRVM} dataset simulation parameters.}
\label{tab:metarvm-params}
\resizebox{\columnwidth}{!}{
\begin{tabular}{c|c|c}
\toprule
\textbf{Input} & \textbf{Meaning} & \textbf{Bound} \\
\midrule
$ts$ &transmissibility for susceptible& (0.1, 0.9) \\
$tv$ &transmissibility for vaccinated& (0.1, 0.9) \\
$dv$ &mean duration in vaccinated state& (30, 90) \\
$de$ &mean duration in exposed state& (1,  5) \\
$dp$ &mean duration in infectious presymptomatic state& (1, 3) \\
$da$ &mean duration in infectious asymptomatic state& (1, 9) \\
$ds$ &mean duration in infectious symptomatic state& (1, 9) \\
$dh$ &mean duration in hospitalized state& (1, 5) \\
$dr$ &mean duration in recovered state& (30, 90) \\
$ve$ &vaccine efficacy& (0.3, 0.8) \\
\bottomrule
\end{tabular}
}
\end{table}

\begin{figure}[htbp]
    \centering
    \subfloat[RMSPE.]{\includegraphics[width=0.5\linewidth]{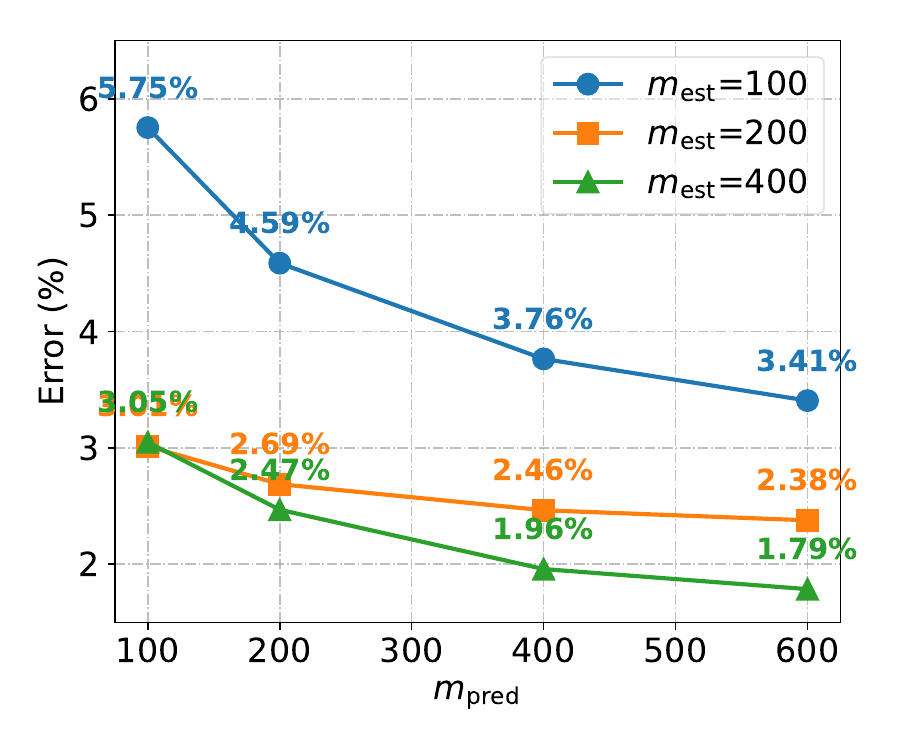}}
    \subfloat[Estimated parameters.]{\includegraphics[width=0.5\linewidth]{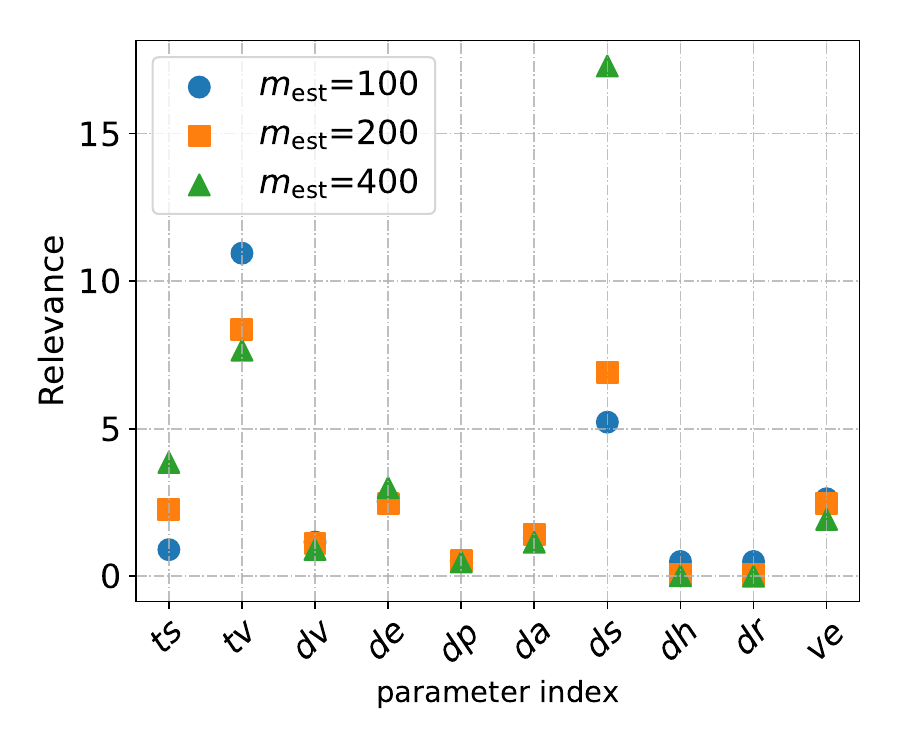}}
    \caption{\textcolor{black}{Root Mean Squared Percentage Error (RMSPE) and estimated relevance} of SBV algorithm on \texttt{MetaRVM} dataset.}
    \label{fig:realmetarvm}
\end{figure}

The results are presented in Figure \ref{fig:realmetarvm}, where a large $m$ for both estimation and prediction generally leads to improved RMSPE. In the parameter estimation, the relevance $dh$ and $dr$ is close to 0, which aligns with our expectation, as these parameters are not involved in the accumulated number of hospitalizations in the \texttt{MetaRVM} simulator. Besides, for $ds$ and $tv$, the estimation varies significantly to the increasing value of $m_{\text{est}}$.

\vspace{-3mm}
\section{Performance and Energy Evaluation}
\begin{figure*}[htbp]
    \centering

    \begin{minipage}[c]{0.23\linewidth}
        \centering
        \includegraphics[width=\linewidth]{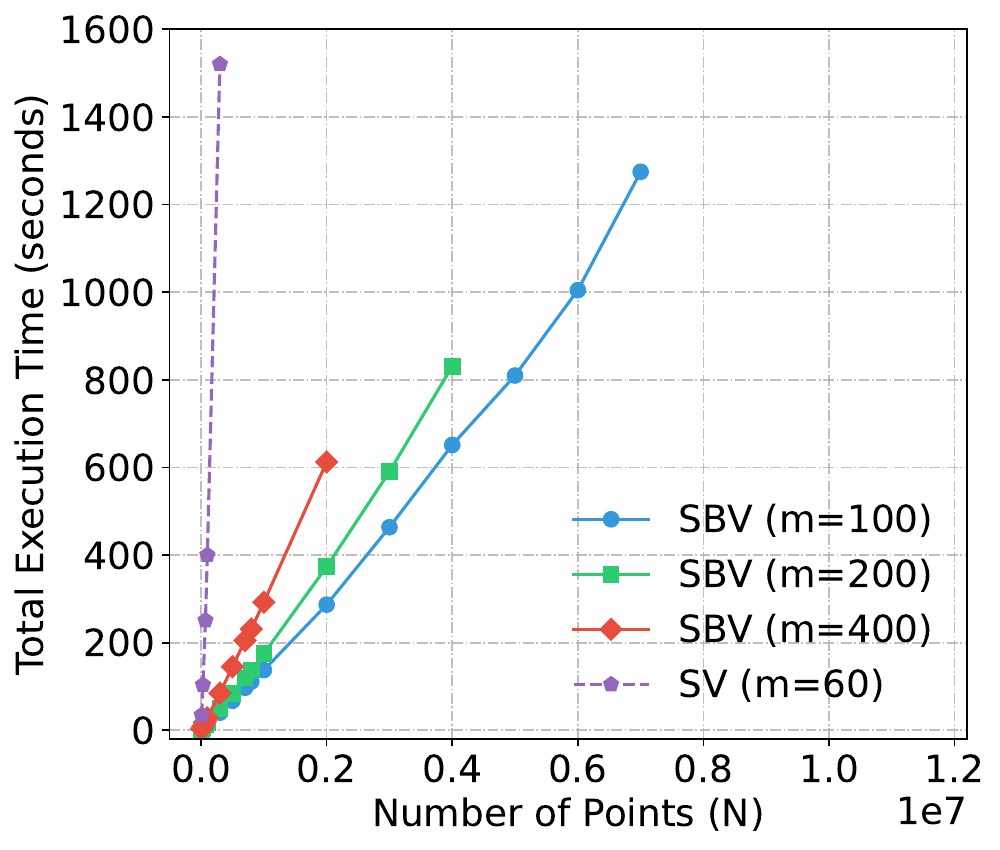}\\
        \vspace{2mm}
        \small \textbf{(a) Runtime on AMD EPYC/ NVIDIA A100 GPU (CPU+GPU).}
    \end{minipage}
    \hfill
    \begin{minipage}[c]{0.23\linewidth}
        \centering
        \includegraphics[width=\linewidth]{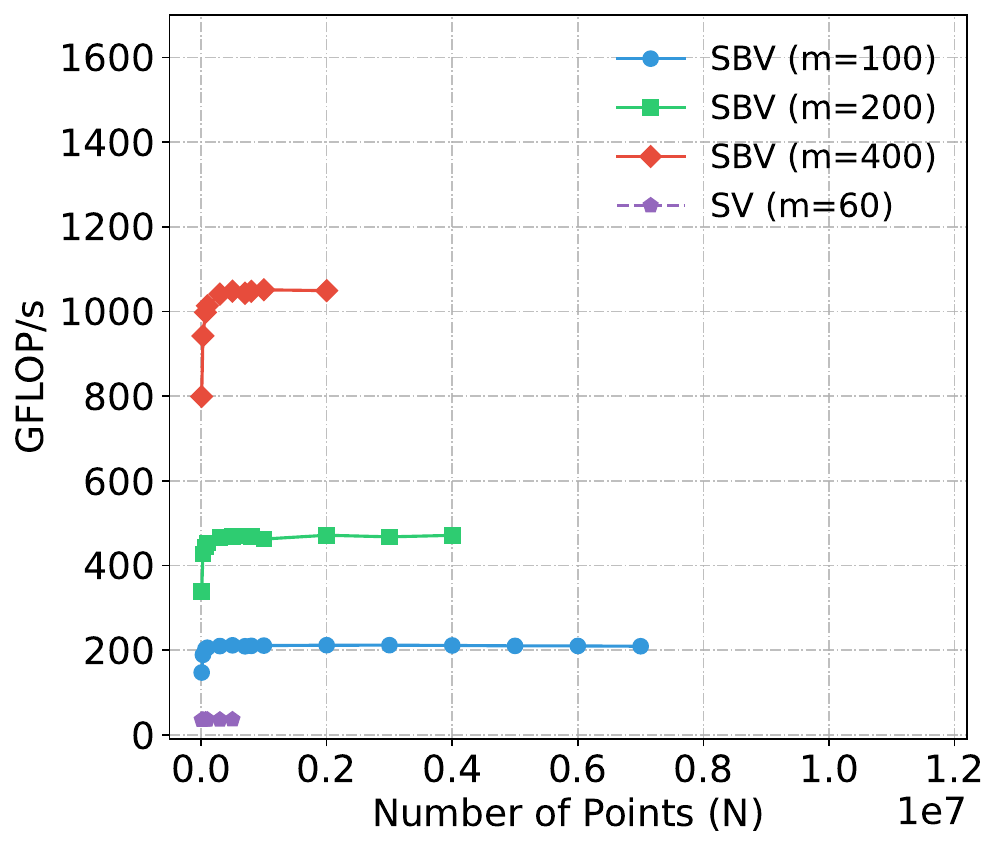}\\
        \vspace{2mm}
        \small \textbf{(b) Performance on AMD EPYC/ NVIDIA A100 GPU  (GPU only).}
    \end{minipage}
    \hfill
    \begin{minipage}[c]{0.23\linewidth}
        \centering
        \includegraphics[width=\linewidth]{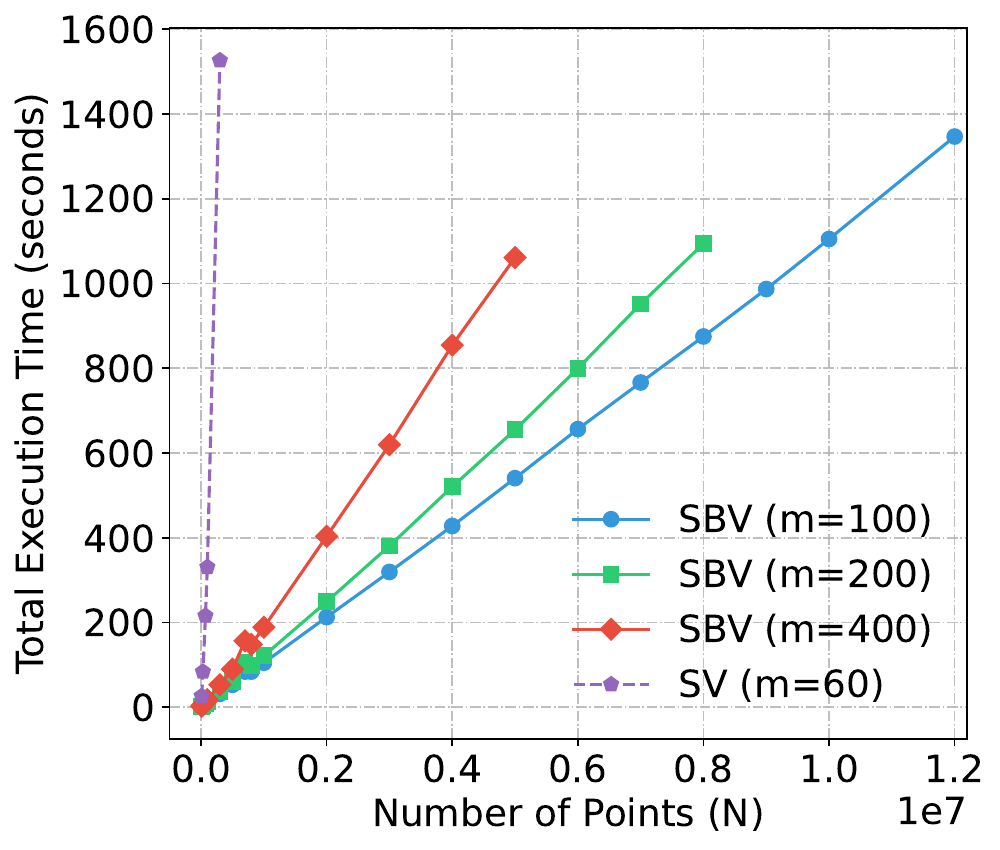}\\
        \vspace{2mm}
        \small \textbf{(c) Runtime on  GH200 superchip\\ (CPU+GPU).}
    \end{minipage}
    \hfill
    \begin{minipage}[c]{0.23\linewidth}
        \centering
        \includegraphics[width=\linewidth]{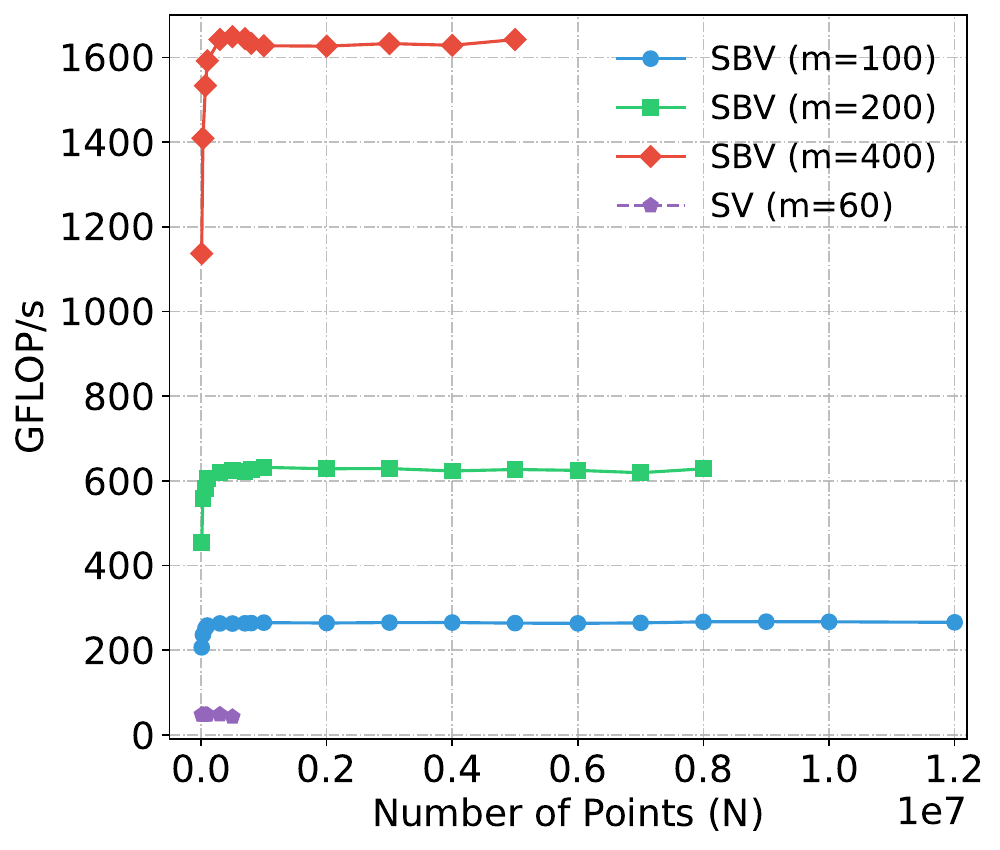}\\
        \vspace{2mm}
        \small \textbf{(d) Performance on GH200 superchip\\  (GPU only).}
    \end{minipage}

    \caption{Performance comparison of SBV and SV methods with 500 MLE iterations on single AMD EPYC with NVIDIA A100 (40~GB) and single NVIDIA GH200 superchip (96~GB). 
    }
    \label{fig:perfsinglegpu}
\end{figure*}


We conduct experiments on two modern NVIDIA GPU architectures, A100 and GH200, hosted at the Jülich Supercomputing Centre (JSC) on the JURECA-DC and JUPITER systems including the early access JUPITER development system JEDI. JUPITER features about 6000 nodes, each equipped with four NVIDIA GH200 Grace Hopper Superchips. 48 of these nodes were accessible early as the JEDI system. Each Superchip integrates a 72-core Grace CPU (3.1 GHz, 120 GB) and a Hopper GPU (96 GB HBM3), connected via NVLink-C2C (900 GB/s), with a total power consumption of approximately 680 W per Superchip. JURECA-DC contains 192 GPU-equipped nodes, each with two 64-core AMD EPYC 7742 CPUs (2.25 GHz, 512 GB, 225 W per socket) and four NVIDIA A100 GPUs (40 GB HBM2, 400 W per GPU). Both systems use InfiniBand interconnects: JUPITER and JEDI provide four NDR200 (200 Gbit/s) links per node, while JURECA-DC GPU offers two HDR (200 Gbit/s) links per node. Our framework is compiled with GCC 13.3.0, uses CUDA 11.8/12.0, and links against MAGMA 2.7.2~\cite{abdelfattah2021set, agullo2009numerical} and NLopt 2.7.1. \textcolor{black}{All experiments run in double precision and are repeated for consistency.}

\textcolor{black}{
All following experiments set the default $bs=100$, 10-dimensional input $[0, 1]^{10}$, and only consider the estimation stage, which is the most time-consuming part in the whole lifetime of emulations; thus, $m$ represents the number of nearest neighbors for estimations and uses Maximum Likelihood Estimation (MLE). Here, we set the number of iterations to 500, as the average number of optimization rounds is 500 on benchmark datasets.
}


\subsection{Single-Node Performance}


Figure~\ref{fig:perfsinglegpu} illustrates the runtime and GPU throughput performance of the SBV and SV methods on two nodes: an NVIDIA A100 with its host CPU, and the GH200 Superchip. Subfigures~(a) and (c) show that SBV consistently outperforms SV in total execution time across different values of $m$, demonstrating superior scalability as the dataset size increases. As expected, larger values of $m$ increase the total runtime due to the higher computational cost per iteration; however, they also yield significantly better approximation and prediction accuracy, as shown in the accuracy section.

Subfigures (b) and (d) highlight the GPU throughput (in GFLOP/s) for a single iteration, showing that SBV achieves significantly higher sustained performance than SV on both platforms. Notably, the GH200 delivers \textcolor{black}{performance} exceeding $1600$ GFLOP/s with $m = 400$. 
These results confirm that SBV is faster and more GPU-efficient than SV, particularly when utilizing larger neighborhoods on modern architectures, such as the GH200. 

\textcolor{red}{

While the GH200 offers substantially higher theoretical FP64 peak performance compared to the A100, the SBV workload is not exclusively compute-bound. The primary kernels involve batched Cholesky factorizations and moderate-sized GEMM/TRSM operations on blocks with $bs \approx 100$ and $m \leq 400$. For these matrix dimensions, arithmetic intensity remains moderate, and performance is limited in part by memory bandwidth and batched-kernel occupancy rather than by peak FP64 throughput. As a result, the observed performance gains correspond more closely to increases in effective HBM bandwidth than to theoretical peak FLOP ratios. This pattern suggests that SBV operates within a mixed memory-and-compute regime, rather than failing to utilize the capabilities of the newer architecture.
}

\subsection{Scalability Across Multiple Nodes}

\begin{figure*}
    \centering
    \subfloat[Weak scalability and PE (A100) (2M–1024M points).]{\includegraphics[width=0.47\linewidth]{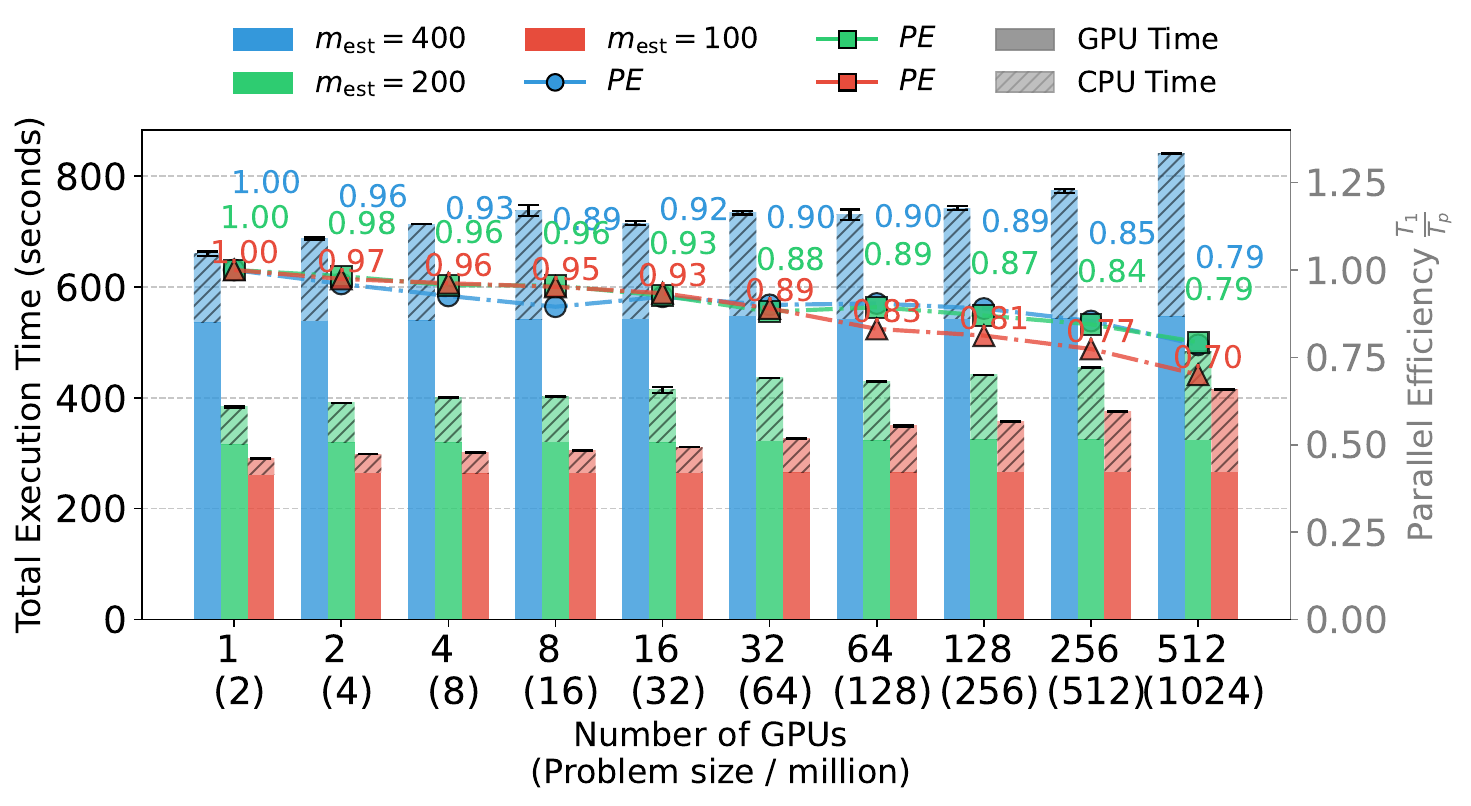}}
    \subfloat[Strong Scalability and PE (A100) (2M points).] {\includegraphics[width=0.47\linewidth]{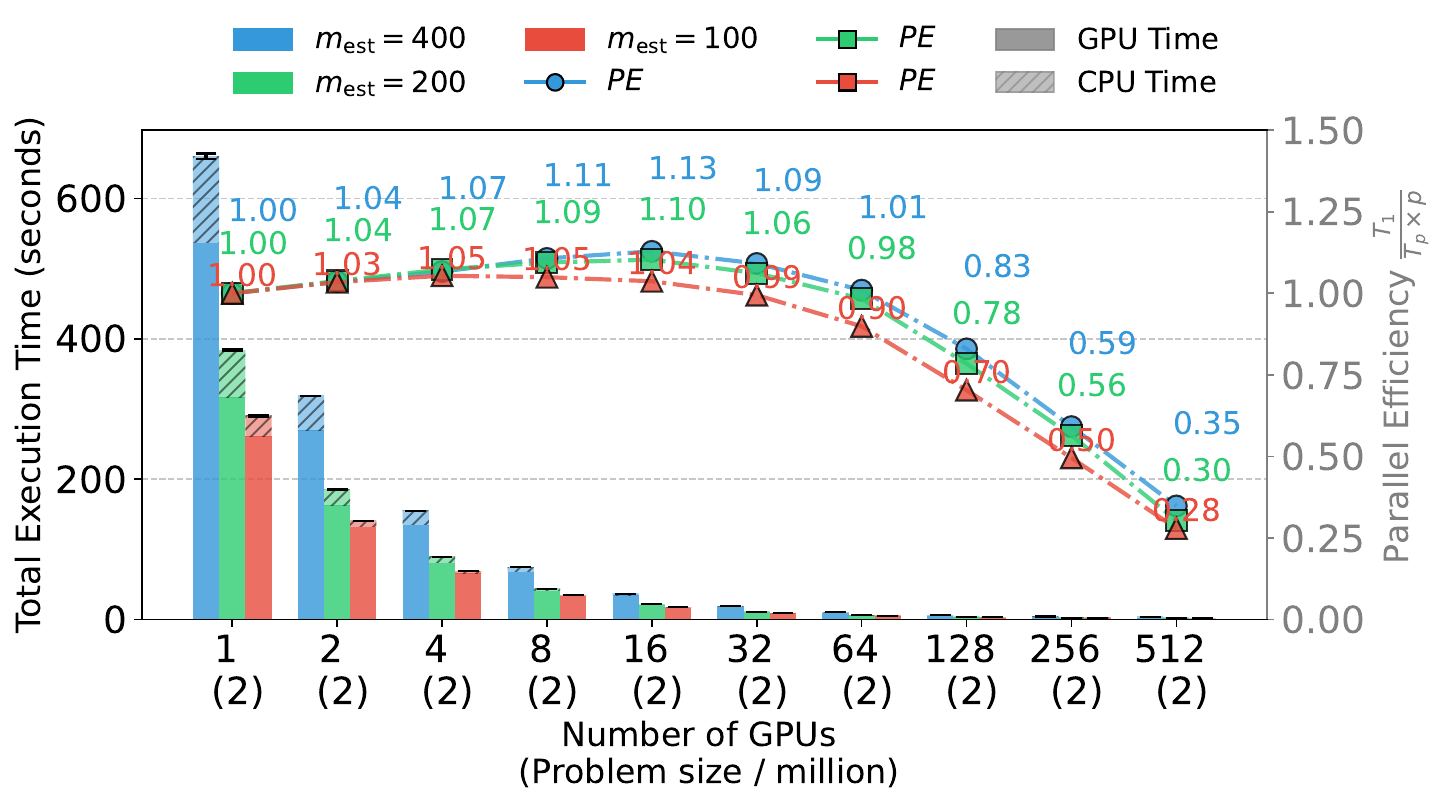}}
     
    \subfloat[Weak scalability and PE (GH200) (5M–2.56B points).]{\includegraphics[width=0.48\linewidth]{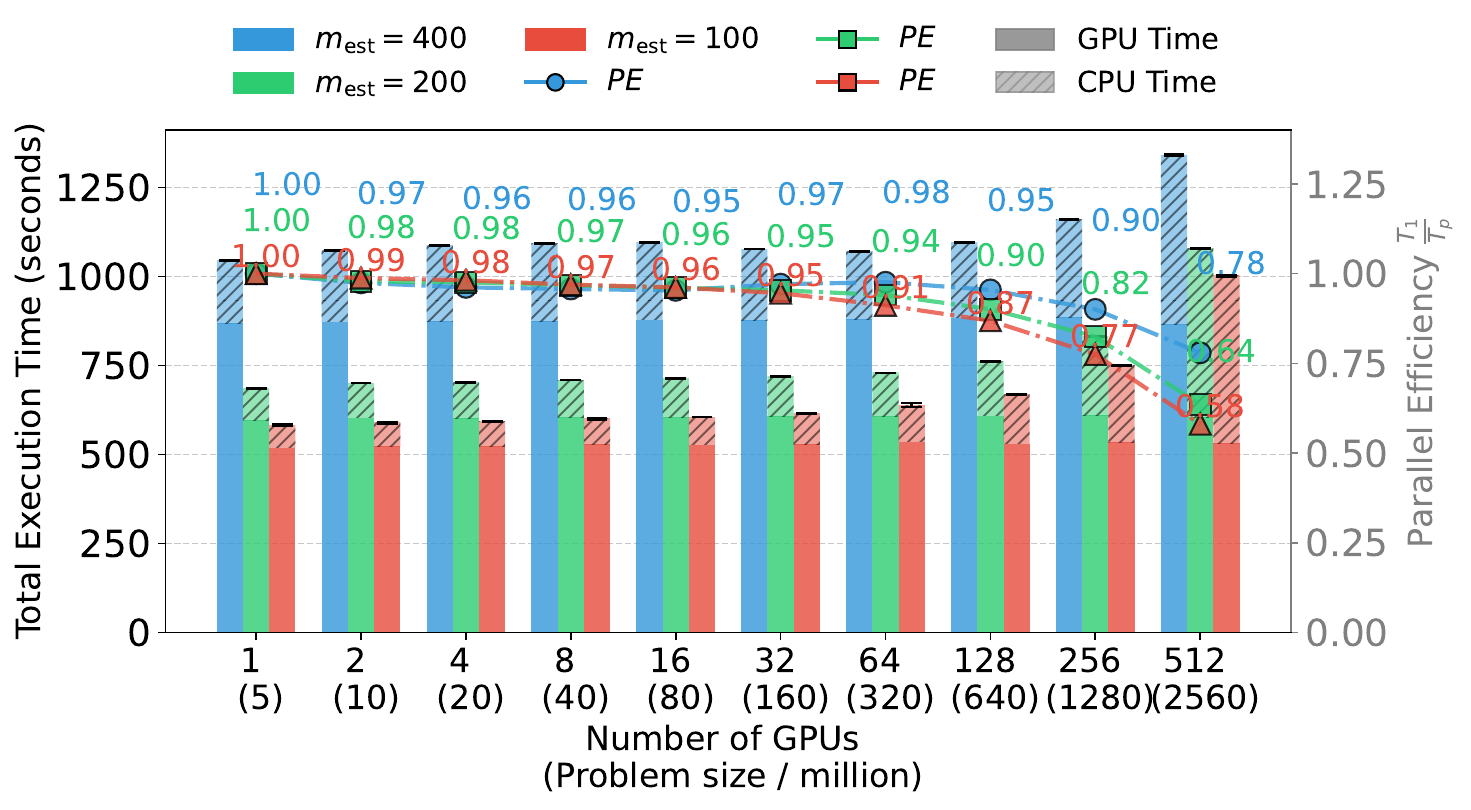}}
    \subfloat[Strong Scalability and PE (GH200) (5M points).]{\includegraphics[width=0.48\linewidth]{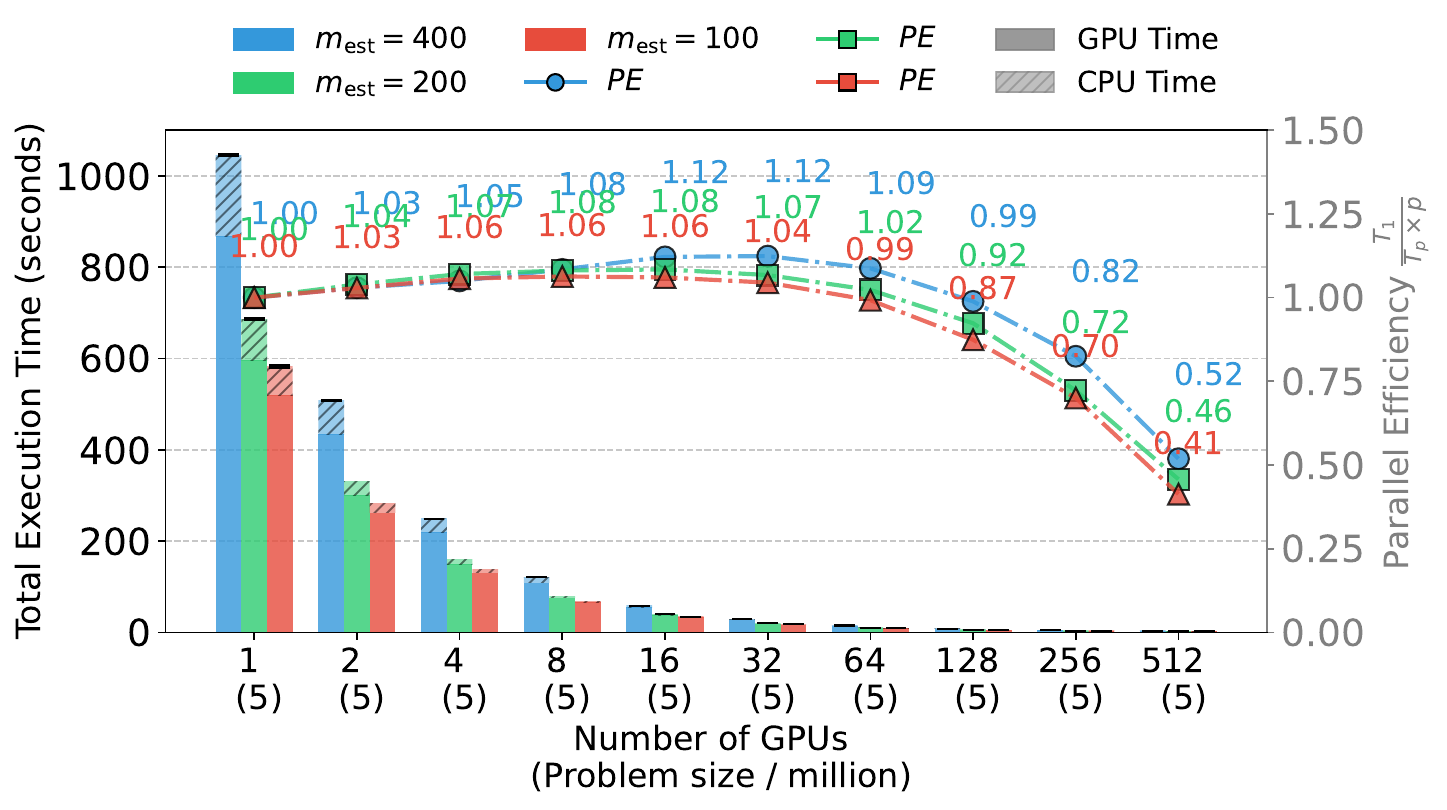}}
\caption{Weak and strong scaling of SBV were evaluated on up to 512 GPUs, using 500 MLE iterations on two architectures: AMD EPYC with NVIDIA A100 (40 GB, up to 128M points) and the NVIDIA GH200 superchip (96 GB, up to 2.56B points); \textcolor{red}{$T_1$ and $T_p$ represent the consumed time for single and $p$ GPUs, respectively, in the Parallel Efficiency (PE) = $\frac{T_1}{n \, T_p}$.}\vspace{-6mm}}

    \label{fig:scaling}
\end{figure*}

Figure~\ref{fig:scaling} demonstrates the weak and strong scaling performance of the distributed SBV implementation on NVIDIA A100 and GH200 architectures, scaling up to 512 GPUs. 
In weak scaling (Subfigures (a) and (c)), both systems maintain high parallel efficiency ($PE$) as the problem size increases proportionally with the number of GPUs. \textcolor{black}{
The drops in the larger scale are derived from the CPU computation, specifically Step 2 in the Algorithm \ref{alg:nn}, which can be reduced using a smaller $\lambda$ in Eq~\ref{eq:distancethreashold}. Specifically, the candidate count does not scale proportionally and retains some stochastic variability, introducing non-uniform load balancing across tasks. A gradually decreasing $\lambda$ is recommended in the NNS as the number of GPUs increases.}

In strong scaling (Subfigures (b) and (d)), total execution time decreases nearly linearly with increasing GPU count, and parallel efficiency exceeds ideal scaling ($PE > 1.0$) in several cases due to \textcolor{black}{less nearest neighbor candidates in the Algorithm 4, where $\lambda$ can be increased accordingly to ensure there is enough candidate for NNS. The drops in the larger GPU count originate from the overhead of multiple function calls, where the 500 iterations are executed, and their accumulation cannot be hidden as the computational task is light for each GPU, where each iteration only costs around 30 milliseconds.}

In both scaling experiments, the influence of the neighborhood size $m_{\text{est}}$ is also evident: larger values, such as $m = 400$, increase the computational workload per point and runtime, but yield higher GPU throughput and overall hardware utilization. Smaller values, such as $m = 100$, result in faster runtimes but less efficient GPU utilization, especially on the GH200. Mid-range values (e.g., $m = 200$) balance performance and efficiency. 

A detailed breakdown shows that GPU compute time dominates execution, accounting for over 80\% of the total runtime in large problem sizes. However, the remaining CPU-side computation introduces scaling variations. On A100, limited CPU resources (32 cores per task) lead to a gradual drop in \(PE\), whereas the 72-core CPU on GH200 better hides this overhead. This difference arises from the CPU's role in maintaining a globally sorted array for block reordering and candidate filtering, as in Equation~\eqref{eq:mmsearchingBV}.


\begin{figure*}

\end{figure*}


\subsection{Power Consumption Analysis}

\begin{figure*}[htbp]
    \centering
    \subfloat[Single NVIDIA A100 GPU  (2M points).\label{fig:power_a100}]
    {
        \includegraphics[width=0.34\linewidth]{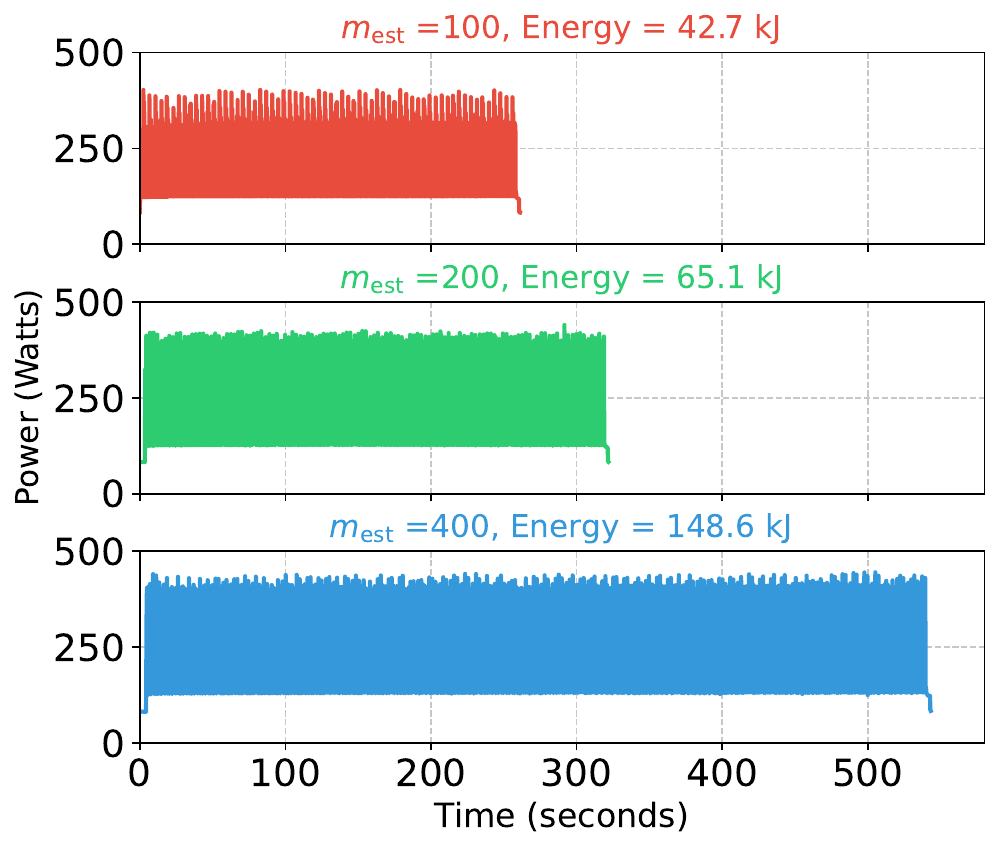}
    }
    \hspace{15mm}
    \subfloat[Single NVIDIA H100 GPU (5M points).\label{fig:power_gh200}]
    {
        \includegraphics[width=0.34\linewidth]{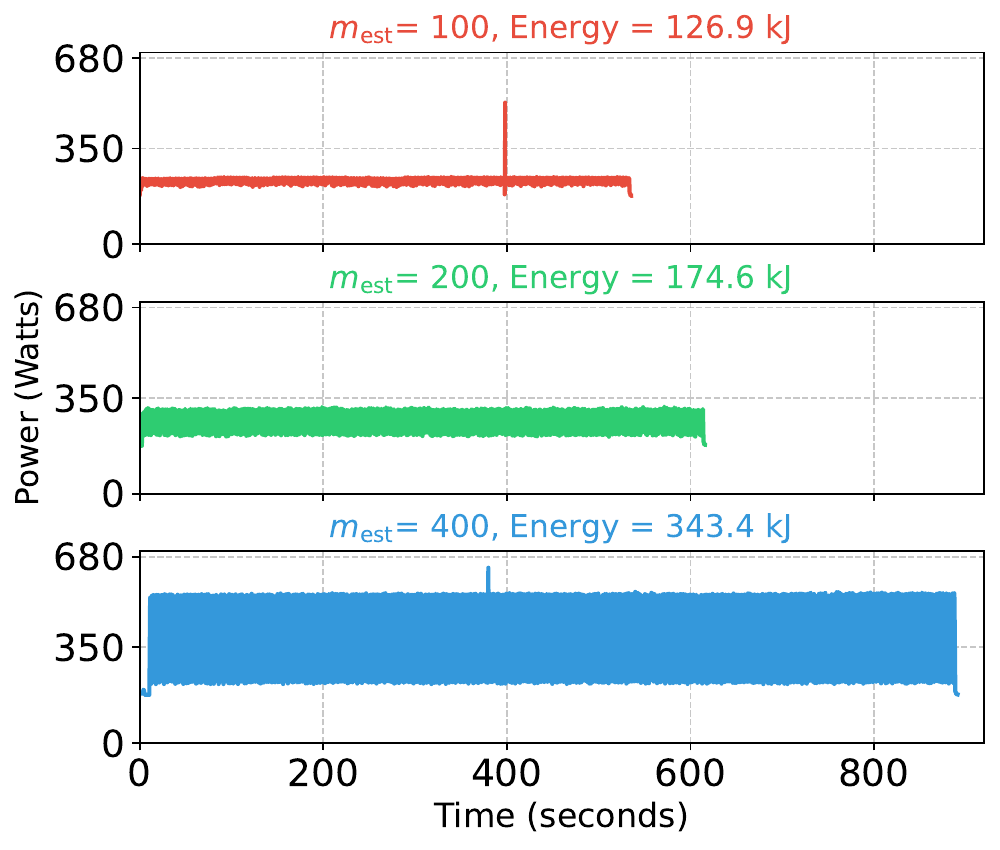}
    }
    \caption{Power consumption/energy (kJ) on a single GPU for two NVIDIA GPUs over 500 iterations per $m_{\text{est}}$.  A100 GPU with 400 W max power cap and Hopper GPU with (680 W - the power usage of the Grace CPU, RAM, and system I/O) max power cap. \vspace{-4mm}}
    \label{fig:power}
\end{figure*}

Figure~\ref{fig:power} presents the power consumption profiles of a single A100 and GH200 superchip GPU while running 500 log-likelihood iterations across varying neighborhood sizes ($m_{\text{est}}$), using a problem size of 2M points on A100 and 5M on GH200, the largest size that fits in GPU memory. As expected, both power draw and total energy consumption increase with larger $m_{\text{est}}$ due to the higher computational workload. On A100 (Figure~\ref{fig:power_a100}), power remains well below the 500 W limit, with total energy ranging from 42.7 kJ ($m_{\text{est}}=100$) to 148.6 kJ ($m_{\text{est}}=400$). In contrast, the Hopper GPUs as part of the GH200 superchips (Figure~\ref{fig:power_gh200}) consume significantly more energy, up to 343.4 kj, while handling a larger problem size, enabled by a higher power envelope. For GH200 superchips, the maximum power consumption is per superchip (680 W for each GH200 in JEDI) and the Grace CPU is prioritized. To avoid power shortage and thus performance drops of the Hopper GPUs, the maximum CPU power consumption can be capped (including RAM and system I/O). On the JEDI system, the default CPU power cap is 100 W, leaving at least 580 W for the Hopper GPUs. 

We also compare our results to the power profiles of exact GPs reported in~\cite{cao2023reducing}, which measure the energy consumption of Cholesky-based FP64 and exact GP variants on V100, A100, and H100 GPUs. In that study, a single MLE iteration requires more than $140$ kJ on A100 and $340$ kJ on H100, even for relatively small matrix sizes ($122{,}880$ points). In contrast, our full SBV-based MLE estimation, which includes 500 iterations on much larger datasets (2M and 5M points), consumes only $\sim$12–40\% of the energy required for a single exact GP iteration on those systems, despite handling problems that are 16$\times$ to 40$\times$ larger. These results highlight the energy efficiency of SBV on modern GPU architectures while demonstrating its scalability for high-dimensional GP modeling.

\section{Conclusion}

This paper introduced the Scaled Block Vecchia (SBV) approximation, a novel framework for scalable GP emulation tailored for high-dimensional datasets on distributed GPU systems. By integrating anisotropic input scaling with block-based conditioning and leveraging batched GPU kernels, SBV substantially reduces memory and computational costs while maintaining high predictive accuracy. To our knowledge, this is the first distributed implementation of a Vecchia-based GP approximation. Through extensive experiments on synthetic and real-world data, including the satellite drag benchmark and the MetaRVM epidemiological model, we demonstrate scalability up to $512$ GPUs and the ability to handle up to $2.5$B points, achieving energy efficiency and state-of-the-art accuracy.



\section*{Acknowledgments}
This work was supported by King Abdullah University of Science and Technology (KAUST). We gratefully acknowledge Ahmad Abdelfattah from the University of Tennessee, Knoxville (UTK), for his support and assistance in optimizing the use of the MAGMA library. We also thank the KAUST Supercomputing Laboratory (KSL) team for providing access to the Ibex cluster for initial runs and GPU experiments. We further acknowledge the computing time provided by the John von Neumann Institute for Computing (NIC) on the supercomputer JURECA-DC. This project received access to JEDI, a preparation system for the JUPITER supercomputer, and JUPITER through the JUPITER Research and Early Access Program (JUREAP). JUPITER is funded by the EuroHPC Joint Undertaking, the German Federal Ministry of Research, Technology, and Space, and the Ministry of Culture and Science of the German state of North Rhine-Westphalia.

\balance





\bibliographystyle{ACM-Reference-Format}
\bibliography{sample-base}

@String{Computing = "Computing" }

@String{Computer = "{IEEE} Computer" }

@String{Springer = "Springer-Verlag" }

@article{winkel2021sequential,
  title={Sequential optimization in locally important dimensions},
  author={Winkel, Munir A and Stallrich, Jonathan W and Storlie, Curtis B and Reich, Brian J},
  journal={Technometrics},
  volume={63},
  number={2},
  pages={236--248},
  year={2021},
  publisher={Taylor \& Francis}
}

@article{saltelli2010variance,
  title={Variance based sensitivity analysis of model output. Design and estimator for the total sensitivity index},
  author={Saltelli, Andrea and Annoni, Paola and Azzini, Ivano and Campolongo, Francesca and Ratto, Marco and Tarantola, Stefano},
  journal={Computer Physics Communications},
  volume={181},
  number={2},
  pages={259--270},
  year={2010},
  publisher={Elsevier}
}

@article{NNsurvey,
  title={A survey on nearest neighbor search methods},
  author={Abbasifard, Mohammad Reza and Ghahremani, Bijan and Naderi, Hassan},
  journal={International Journal of Computer Applications},
  volume={95},
  number={25},
  year={2014},
  publisher={Citeseer}
}

@inproceedings{NNconstraint,
  title={Constrained nearest neighbor queries},
  author={Ferhatosmanoglu, Hakan and Stanoi, Ioanna and Agrawal, Divyakant and El Abbadi, Amr},
  booktitle={International Symposium on Spatial and Temporal Databases},
  pages={257--276},
  year={2001},
  organization={Springer}
}

@article{vecchia1988estimation,
  title={Estimation and model identification for continuous spatial processes},
  author={Vecchia, Aldo V},
  journal={Journal of the Royal Statistical Society Series B: Statistical Methodology},
  volume={50},
  number={2},
  pages={297--312},
  year={1988},
  publisher={Oxford University Press}
}

@article{clarkson2006nearest,
  title={Nearest-neighbor searching and metric space dimensions},
  author={Clarkson, Kenneth L and others},
  journal={Nearest-neighbor methods for learning and vision: theory and practice},
  pages={15--59},
  year={2006}
}

@article{katzfuss2022scaled,
  title={Scaled Vecchia approximation for fast computer-model emulation},
  author={Katzfuss, Matthias and Guinness, Joseph and Lawrence, Earl},
  journal={SIAM/ASA Journal on Uncertainty Quantification},
  volume={10},
  number={2},
  pages={537--554},
  year={2022},
  publisher={SIAM}
}

@article{genton2001classes,
  title={Classes of kernels for machine learning: a statistics perspective},
  author={Genton, Marc G},
  journal={Journal of Machine Learning Research},
  volume={2},
  number={Dec},
  pages={299--312},
  year={2001}
}

@article{sun2019emulating,
  title={Emulating satellite drag from large simulation experiments},
  author={Sun, Furong and Gramacy, Robert B and Haaland, Benjamin and Lawrence, Earl and Walker, Andrew},
  journal={SIAM/ASA Journal on Uncertainty Quantification},
  volume={7},
  number={2},
  pages={720--759},
  year={2019},
  publisher={SIAM}
}

@article{abdulah2021accelerating,
  title={Accelerating geostatistical modeling and prediction with mixed-precision computations: A high-productivity approach with PaRSEC},
  author={Abdulah, Sameh and Cao, Qinglei and Pei, Yu and Bosilca, George and Dongarra, Jack and Genton, Marc G and Keyes, David E and Ltaief, Hatem and Sun, Ying},
  journal={IEEE Transactions on Parallel and Distributed Systems},
  volume={33},
  number={4},
  pages={964--976},
  year={2021},
  publisher={IEEE}
}

@inproceedings{salvana2022parallel,
  title={Parallel space-time likelihood optimization for air pollution prediction on large-scale systems},
  author={Salva{\~n}a, Mary Lai O and Abdulah, Sameh and Ltaief, Hatem and Sun, Ying and Genton, Marc G and Keyes, David E},
  booktitle={Proceedings of the Platform for Advanced Scientific Computing Conference},
  pages={1--11},
  year={2022}
}

@article{katzfuss2021general,
  title={A general framework for Vecchia approximations of Gaussian processes},
  author={Katzfuss, Matthias and Guinness, Joseph},
  journal={Statistical Science},
  volume={36},
  number={1},
  pages={124--141},
  year={2021},
  publisher={JSTOR}
}

@article{mondal2023tile,
  title={Tile low-rank approximations of non-Gaussian space and space-time Tukey g-and-h random field likelihoods and predictions on large-scale systems},
  author={Mondal, Sagnik and Abdulah, Sameh and Ltaief, Hatem and Sun, Ying and Genton, Marc G and Keyes, David E},
  journal={Journal of Parallel and Distributed Computing},
  volume={180},
  pages={104715},
  year={2023},
  publisher={Elsevier}
}

@article{shi2025decentralized,
  title={Decentralized Inference for Distributed Geospatial Data Using Low-Rank Models},
  author={Shi, Jianwei and Abdulah, Sameh and Sun, Ying and Genton, Marc G},
  journal={arXiv preprint arXiv:2502.00309},
  year={2025}
}

@article{eyring2016overview,
  title={Overview of the Coupled Model Intercomparison Project Phase 6 (CMIP6) experimental design and organization},
  author={Eyring, Veronika and Bony, Sandrine and Meehl, Gerald A and Senior, Catherine A and Stevens, Bjorn and Stouffer, Ronald J and Taylor, Karl E},
  journal={Geoscientific Model Development},
  volume={9},
  number={5},
  pages={1937--1958},
  year={2016},
  publisher={Copernicus GmbH}
}

@article{arora2023review,
  title={A review of radial basis function with applications explored},
  author={Arora, Geeta and Bala, Kiran and Emadifar, Homan and Khademi, Masoumeh},
  journal={Journal of the Egyptian Mathematical Society},
  volume={31},
  number={1},
  pages={1--14},
  year={2023},
  publisher={National Information and Documentation Centre (NIDOC), Academy of Scientific~…}
}

@article{kundig2025iterative,
  title={Iterative methods for {V}ecchia-{L}aplace approximations for latent gaussian process models},
  author={K{\"u}ndig, Pascal and Sigrist, Fabio},
  journal={Journal of the American Statistical Association},
  volume={120},
  number={550},
  pages={1267--1280},
  year={2025},
  publisher={Taylor \& Francis}
}

@article{cao2022scalable,
  title={Scalable Gaussian-process regression and variable selection using Vecchia approximations},
  author={Cao, Jian and Guinness, Joseph and Genton, Marc G and Katzfuss, Matthias},
  journal={Journal of machine learning research},
  volume={23},
  number={348},
  pages={1--30},
  year={2022}
}

@inproceedings{jimenez2023scalable,
  title={Scalable bayesian optimization using vecchia approximations of gaussian processes},
  author={Jimenez, Felix and Katzfuss, Matthias},
  booktitle={International Conference on Artificial Intelligence and Statistics},
  pages={1492--1512},
  year={2023},
  organization={PMLR}
}

@article{huser2024vecchia,
  title={Vecchia likelihood approximation for accurate and fast inference with intractable spatial max-stable models},
  author={Huser, Rapha{\"e}l and Stein, Michael L and Zhong, Peng},
  journal={Journal of Computational and Graphical Statistics},
  volume={33},
  number={3},
  pages={978--990},
  year={2024},
  publisher={Taylor \& Francis}
}

@article{cao2025linear,
  title={Linear-cost Vecchia approximation of multivariate normal probabilities},
  author={Cao, Jian and Katzfuss, Matthias},
  journal={Journal of the American Statistical Association},
  pages={1--14},
  year={2025},
  publisher={Taylor \& Francis}
}

@article{sauer2023vecchia,
  title={Vecchia-approximated deep Gaussian processes for computer experiments},
  author={Sauer, Annie and Cooper, Andrew and Gramacy, Robert B},
  journal={Journal of Computational and Graphical Statistics},
  volume={32},
  number={3},
  pages={824--837},
  year={2023},
  publisher={Taylor \& Francis}
}

@article{acosta1995radial,
  title={Radial basis function and related models: an overview},
  author={Acosta, Felipe Miguel Aparicio},
  journal={Signal Processing},
  volume={45},
  number={1},
  pages={37--58},
  year={1995},
  publisher={Elsevier}
}

@article{walters2018bayesian,
  title={Bayesian calibration of strength parameters using hydrocode simulations of symmetric impact shock experiments of Al-5083},
  author={Walters, David J and Biswas, Ayan and Lawrence, Earl C and Francom, Devin C and Luscher, Darby J and Fredenburg, D Anthony and Moran, Kelly R and Sweeney, Christine M and Sandberg, Richard L and Ahrens, James P and others},
  journal={Journal of Applied Physics},
  volume={124},
  number={20},
  year={2018},
  publisher={AIP Publishing}
}

@article{springel2005cosmological,
  title={The cosmological simulation code GADGET-2},
  author={Springel, Volker},
  journal={Monthly Notices of the Royal Astronomical Society},
  volume={364},
  number={4},
  pages={1105--1134},
  year={2005},
  publisher={The Royal Astronomical Society}
}

@article{spurio2022cosmopower,
  title={CosmoPower: emulating cosmological power spectra for accelerated Bayesian inference from next-generation surveys},
  author={Spurio Mancini, Alessio and Piras, Davide and Alsing, Justin and Joachimi, Benjamin and Hobson, Michael P},
  journal={Monthly Notices of the Royal Astronomical Society},
  volume={511},
  number={2},
  pages={1771--1788},
  year={2022},
  publisher={Oxford University Press}
}

@article{borvik2001computational,
  title={A computational model of viscoplasticity and ductile damage for impact and penetration},
  author={B{\o}rvik, T and Hopperstad, OS and Berstad, T and Langseth, M},
  journal={European Journal of Mechanics-A/Solids},
  volume={20},
  number={5},
  pages={685--712},
  year={2001},
  publisher={Elsevier}
}

@article{kochkov2024neural,
  title={Neural general circulation models for weather and climate},
  author={Kochkov, Dmitrii and Yuval, Janni and Langmore, Ian and Norgaard, Peter and Smith, Jamie and Mooers, Griffin and Kl{\"o}wer, Milan and Lottes, James and Rasp, Stephan and D{\"u}ben, Peter and others},
  journal={Nature},
  volume={632},
  number={8027},
  pages={1060--1066},
  year={2024},
  publisher={Nature Publishing Group UK London}
}

@book{rasmussen2006gaussian,
  title={Gaussian Processes for Machine Learning},
  author={Rasmussen, Carl Edward and Williams, Christopher KI},
  year={2006},
  publisher={MIT Press}
}

@inproceedings{cao2023reducing,
  title={Reducing data motion and energy consumption of geospatial modeling applications using automated precision conversion},
  author={Cao, Qinglei and Abdulah, Sameh and Ltaief, Hatem and Genton, Marc G and Keyes, David and Bosilca, George},
  booktitle={2023 IEEE International Conference on Cluster Computing (CLUSTER)},
  pages={330--342},
  year={2023},
  organization={IEEE}
}

@article{fadikar2018calibrating,
  title={Calibrating a stochastic, agent-based model using quantile-based emulation},
  author={Fadikar, Arindam and Higdon, Dave and Chen, Jiangzhuo and Lewis, Bryan and Venkatramanan, Srinivasan and Marathe, Madhav},
  journal={SIAM/ASA Journal on Uncertainty Quantification},
  volume={6},
  number={4},
  pages={1685--1706},
  year={2018},
  publisher={SIAM}
}

@book{cover2006elements,
  title={Elements of Information Theory},
  author={Cover, Thomas M and Thomas, Joy A},
  year={2006},
  publisher={Wiley-Interscience}
}

@book{murphy2012machine,
  title={Machine Learning: A Probabilistic Perspective},
  author={Murphy, Kevin P},
  year={2012},
  publisher={MIT Press}
}

@article{lawrence2017mira,
  title={The mira-titan universe. II. Matter power spectrum emulation},
  author={Lawrence, Earl and Heitmann, Katrin and Kwan, Juliana and Upadhye, Amol and Bingham, Derek and Habib, Salman and Higdon, David and Pope, Adrian and Finkel, Hal and Frontiere, Nicholas},
  journal={The Astrophysical Journal},
  volume={847},
  number={1},
  pages={50},
  year={2017},
  publisher={IOP Publishing}
}

@article{abdelfattah2021set,
  title={A set of batched basic linear algebra subprograms and LAPACK routines},
  author={Abdelfattah, Ahmad and Costa, Timothy and Dongarra, Jack and Gates, Mark and Haidar, Azzam and Hammarling, Sven and Higham, Nicholas J and Kurzak, Jakub and Luszczek, Piotr and Tomov, Stanimire and others},
  journal={ACM Transactions on Mathematical Software (TOMS)},
  volume={47},
  number={3},
  pages={1--23},
  year={2021},
  publisher={ACM New York, NY, USA}
}

@inproceedings{agullo2009numerical,
  title={Numerical linear algebra on emerging architectures: The {PLASMA} and {MAGMA} projects},
  author={Agullo, Emmanuel and Demmel, Jim and Dongarra, Jack and Hadri, Bilel and Kurzak, Jakub and Langou, Julien and Ltaief, Hatem and Luszczek, Piotr and Tomov, Stanimire},
  booktitle={Journal of Physics: Conference Series},
  volume={180},
  number={1},
  pages={012037},
  year={2009},
  organization={IOP Publishing}
}

@article{furrer2006covariance,
  title={Covariance tapering for interpolation of large spatial datasets},
  author={Furrer, Reinhard and Genton, Marc G and Nychka, Douglas},
  journal={Journal of Computational and Graphical Statistics},
  volume={15},
  number={3},
  pages={502--523},
  year={2006},
  publisher={Taylor \& Francis}
}

@article{pan2025block,
  title={Block Vecchia Approximation for Scalable and Efficient Gaussian Process Computations},
  author={Pan, Qilong and Abdulah, Sameh and Genton, Marc G and Sun, Ying},
  journal={Technometrics},
  number={just-accepted},
  pages={1--18},
  year={2025},
  publisher={Taylor \& Francis}
}

@article{tomov2009magma,
  title={Magma library},
  author={Tomov, S and Dongarra, J and Volkov, V and Demmel, J},
  journal={Univ. of Tennessee and Univ. of California, Knoxville, TN, and Berkeley, CA},
  year={2009}
}

@article{MetaRVM,
  title={Developing and deploying a use-inspired metapopulation modeling framework for detailed tracking of stratified health outcomes},
  author={Fadikar, Arindam and Stevens, Abby and Rimer, Sara and Martinez-Moyano, Ignacio and Collier, Nicholson and Ozik, Jonathan and Macal, Charles},
  journal={medRxiv},
  pages={2025--05},
  year={2025},
  publisher={Cold Spring Harbor Laboratory Press},
doi={ https://doi.org/10.1101/2025.05.05.25327021}
}

@article{abdulah2018exageostat,
  title={{ExaGeoStat}: A high performance unified software for geostatistics on manycore systems},
  author={Abdulah, Sameh and Ltaief, Hatem and Sun, Ying and Genton, Marc G and Keyes, David E},
  journal={IEEE Transactions on Parallel and Distributed Systems},
  volume={29},
  number={12},
  pages={2771--2784},
  year={2018},
  publisher={IEEE}
}

@article{guinness2018permutation,
  title={Permutation and grouping methods for sharpening Gaussian process approximations},
  author={Guinness, Joseph},
  journal={Technometrics},
  volume={60},
  number={4},
  pages={415--429},
  year={2018},
  publisher={Taylor \& Francis}
}

@inproceedings{pan2024gpu,
  title={{GPU}-accelerated {Vecchia} approximations of {Gaussian} processes for geospatial data using batched matrix computations},
  author={Pan, Qilong and Abdulah, Sameh and Genton, Marc G and Keyes, David E and Ltaief, Hatem and Sun, Ying},
  booktitle={ISC High Performance 2024 Research Paper Proceedings (39th International Conference)},
  pages={1--12},
  year={2024},
  organization={Prometeus GmbH}
}

@inproceedings{abdulah2019geostatistical,
  title={Geostatistical modeling and prediction using mixed precision tile Cholesky factorization},
  author={Abdulah, Sameh and Ltaief, Hatem and Sun, Ying and Genton, Marc G and Keyes, David E},
  booktitle={2019 IEEE 26th International Conference on High Performance Computing, Data, and Analytics (HiPC)},
  pages={152--162},
  year={2019},
  organization={IEEE}
}

@inproceedings{abdulah2018parallel,
  title={Parallel approximation of the maximum likelihood estimation for the prediction of large-scale geostatistics simulations},
  author={Abdulah, Sameh and Ltaief, Hatem and Sun, Ying and Genton, Marc G and Keyes, David E},
  booktitle={ IEEE International Conference on Cluster Computing},
  pages={98--108},
  year={2018},
  organization={IEEE}
}

@article{bevilacqua2016covariance,
  title={Covariance tapering for multivariate {G}aussian random fields estimation},
  author={Bevilacqua, Moreno and Fass{\`o}, Alessandro and Gaetan, Carlo and Porcu, Emilio and Velandia, Daira},
  journal={Statistical Methods \& Applications},
  volume={25},
  pages={21--37},
  year={2016},
  publisher={Springer}
}

@book{williams2006gaussian,
  title={Gaussian processes for machine learning},
  author={Williams, Christopher KI and Rasmussen, Carl Edward},
  volume={2},
  number={3},
  year={2006},
  publisher={MIT press Cambridge, MA}
}

@article{andrianakis2012effect,
  title={The effect of the nugget on Gaussian process emulators of computer models},
  author={Andrianakis, Ioannis and Challenor, Peter G},
  journal={Computational Statistics \& Data Analysis},
  volume={56},
  number={12},
  pages={4215--4228},
  year={2012},
  publisher={Elsevier}
}

@inproceedings{abdulah2024boosting,
  title={Boosting earth system model outputs and saving petabytes in their storage using exascale climate emulators},
  author={Abdulah, Sameh and Baker, Allison H and Bosilca, George and Cao, Qinglei and Castruccio, Stefano and Genton, Marc G and Keyes, David E and Khalid, Zubair and Ltaief, Hatem and Song, Yan and others},
  booktitle={SC24: International Conference for High Performance Computing, Networking, Storage and Analysis},
  pages={1--12},
  year={2024},
  organization={IEEE}
}

@misc{fiore2018road,
  title={On the road to exascale: Advances in High Performance Computing and Simulations—An overview and editorial},
  author={Fiore, Sandro and Bakhouya, Mohamed and Smari, Waleed W},
  journal={Future Generation Computer Systems},
  volume={82},
  pages={450--458},
  year={2018},
  publisher={Elsevier}
}

@book{santner2003design,
  title={The design and analysis of computer experiments},
  author={Santner, Thomas J and Williams, Brian J and Notz, William I and Williams, Brain J},
  volume={1},
  year={2003},
  publisher={Springer}
}

\appendix

\end{document}